\documentclass[journal]{IEEEtran}

\usepackage{graphicx}
\usepackage{amssymb}
\usepackage{amsmath}
\usepackage{bm}
\usepackage{algorithm}
\usepackage{algorithmic}
\usepackage{color}

\DeclareMathOperator{\Tr}{Tr}
\usepackage{cite}
\usepackage{multirow}
\usepackage{url}
\usepackage{subfigure}
\begin{document}
\title{Joint Precoding and Phase Shift Design in Reconfigurable Intelligent Surfaces-Assisted Secret Key Generation}
%
%
%
\author{Tianyu~Lu,~\IEEEmembership{Graduate~Student~Member,~IEEE,}
  Liquan~Chen,~\IEEEmembership{Senior~Member,~IEEE,}
  Junqing~Zhang,~\IEEEmembership{Member,~IEEE,}
  Chen~Chen,~\IEEEmembership{Graduate~Student Member,~IEEE}
  and~Aiqun~Hu,~\IEEEmembership{Senior~Member,~IEEE}
\thanks{This research is supported by National key research and development program of China, 2020YFE0200600. The work of J. Zhang and C. Chen was also supported by the UK EPSRC under grant ID EP/V027697/1. (\textit{Corresponding author: L. Chen})}
\thanks{T. Lu and L. Chen are with the School of Cyber Science and Engineering, Southeast University, Nanjing, 210096, China. (e-mail: effronlu@seu.edu.cn; lqchen@seu.edu.cn)}
\thanks{J. Zhang and C. Chen are with the Department of Electrical Engineering and Electronics,
University of Liverpool, Liverpool, L69 3GJ, United Kingdom. (email: junqing.zhang@liverpool.ac.uk; c.chen77@liverpool.ac.uk)}
\thanks{A. Hu is with the National Mobile Communications
Research Laboratory, Southeast University, Nanjing 210096, China. (e-mail: aqhu@seu.edu.cn)}
\thanks{L. Chen and A. Hu are also with the Purple Mountain Laboratories for Network and Communication Security, Nanjing, 211111, China.}
}

%
%

\markboth{Journal of \LaTeX\ Class Files,~Vol.~X, No.~X, X~X}%
{Shell \MakeLowercase{\textit{et al.}}: Bare Demo of IEEEtran.cls for IEEE Journals}
%



\maketitle

\begin{abstract}
 Key generation is a promising technique to establish symmetric keys between resource-constrained legitimate users. However, key generation suffers from low secret key rate (SKR) in harsh environments where channel randomness is limited. To address the problem, reconfigurable intelligent surfaces (RISs) are introduced to reshape the channels by controlling massive reflecting elements, which can provide more channel diversity. In this paper, we design a channel probing protocol to fully extract the randomness from the cascaded channel, i.e., the channel through reflecting elements. We derive the analytical expressions of SKR and design a water-filling algorithm based on the Karush-Kuhn-Tucker (KKT) conditions to find the upper bound. To find the optimal precoding and phase shift matrices, we propose an algorithm based on the Grassmann manifold optimization methods. The system is evaluated in terms of SKR, bit disagreement rate (BDR) and randomness. Simulation results show that our protocols significantly improve the SKR as compared to existing protocols.
\end{abstract}


%
\IEEEpeerreviewmaketitle

\section{Introduction}
%
%
%
%
\IEEEPARstart{T}{he} advent of the Internet of Things (IoT) and the development of 5G raise concerns for security in communication networks \cite{intro2}. The number of IoT devices will approach 75 billion by 2025~\cite{zhang2020new}. The information security of current communication and computer systems is protected by symmetric encryption and public key cryptography (PKC). The symmetric encryption, e.g., advanced encryption standard (AES), requires the same key at the ends of legitimate users. The PKC relies on computational hardness assumptions such as discrete logarithm problems. However, traditional cryptographic schemes are not applicable for resource-constrained IoT devices with limited storage and computational power. Furthermore, PKC schemes may be cracked by quantum computers in the future~\cite{zhang2020new}.

Key generation from wireless channels is a promising candidate for establishing symmetric keys between two IoT devices, namely Alice and Bob~\cite{zhang2020new}. Benefiting from the temporal variation, channel reciprocity and spatial diversity properties of wireless media, Alice and Bob can agree on a unique secret key from the measurements of wireless channels. 
\begin{itemize}
	\item \textit{Temporal variation}: When wireless channels change dynamically, there are sufficient randomness to be extracted.
	\item \textit{Channel reciprocity}: The reciprocity between uplink and downlink channels enables Alice and Bob to share a common secret key.
	\item \textit{Spatial diversity}: An eavesdropper, Eve, who is half wavelength away from Alice and Bob, cannot derive any information about the secret key. 
\end{itemize}
Exploiting these features, key generation techniques can achieve information-theoretic security.

There are research explorations both from theoretical and experimental aspects, which demonstrate the potential of key generation \cite{Maurer1993,Ahlswede1993,Maurer2003,zhang2016experimental,ali2013eliminating,zhang2018channel,premnath2014secret}. In the seminal papers of Maurer  \cite{Maurer1993} and Ahlswede and Csiszar  \cite{Ahlswede1993} dated back to 1993, the authors laid an information-theoretic foundation for key generation and proved that Alice and Bob can use correlated randomness source to extract secret keys. Later, Maurer and Wolf investigated the secret key generation over unauthenticated public channels \cite{Maurer2003}. There have also been  many works on realizing practical key generation systems, with experimental exploration using WiFi~\cite{zhang2016experimental}, ZigBee~\cite{ali2013eliminating}, LoRa~\cite{zhang2018channel} and Bluetooth~\cite{premnath2014secret}. 

As key generation relies on the channel randomness and correlated channel measurements, it is challenging to operate well in harsh environments. Firstly, the channel variation is limited in static/quasi-static environments, which cannot provide sufficient randomness~\cite{zhang2020new}. Secondly, when received signals experience low signal-to-noise ratio (SNR), the secret key rate (SKR) will be reduced too~\cite{intro14}. Therefore, it is necessary to develop new techniques to improve SKR in harsh environments.

Recently, reconfigurable intelligent surfaces (RISs) have been introduced to address the above problems. A RIS is regarded as an emerging transmission technology to realize the concept of smart radio environments~\cite{intro3}. RISs empower transceivers to control the scattering characteristics of radio waves, e.g., amplitude, delay, and polarization, by using massive cost-effective reflectors \cite{intro4}. Therefore, RISs have been adopted to solve the problem of low entropy for key generation techniques in static environments~\cite{intro6, intro8, intro9}, where the channel remains near-constant in a long coherence time. The authors proposed to randomly configure the phase shift matrix of a RIS to induce artificial randomness and achieve a one-time pad encryption protocol. Moreover, a RIS can relieve the block effect on key generation by constructing a reflected channel. Compared to traditional relay techniques, e.g. amplify-and-forward (AF) and decode-and-forward (DF), RISs are passive and capable of receiving and transmitting data simultaneously~\cite{intro5}. In \cite{intro10, intro11, intro12}, the authors proposed to optimize the phase shift matrix to increase the SNR at the receiver in order to increase the SKR. 

Key generation from fine-grained channel features can significantly improve the SKR \cite{sim6,intro16,intro18,intro19}. Received signal strength indicator (RSSI) is a popular channel parameter for key generation~\cite{Mathur2008,intro23,intro24}, but its coarse-grained nature limits key generation rate. Compared to RSSI, channel state information (CSI) is a fine-grained channel feature that provides more channel information. The authors in \cite{intro16} analyzed the SKR extracted from the channel coefficients of all subcarriers in orthogonal frequency-division multiplexing (OFDM) systems. Later, the authors in \cite{sim6} achieved a practical key generation protocol based on IEEE 802.11 OFDM systems. In addition, the randomness in the spatial domain can also be employed. The authors in \cite{intro18} proposed to exploit the randomness from multiple antennas.  In the RIS-based key generation, there are multiple channels. 
\begin{itemize}
	\item Direct channel is between Alice and Bob consisting line-of-sight (LoS) (when there is) and non-LoS (NLoS) components not involving RIS.
	\item RIS-reflected subchannels: each subchannel refers to a channel from Alice to an RIS element and then to Bob.
\end{itemize}
Existing RIS-empowered key generation works extract key from the equivalent channel, which is combined by the direct channel and RIS-reflected subchannels. The authors in~\cite{intro12} designed the phase shift matrix and modified the reflected channel so that Alice and Bob can extract more SKR from the equivalent channels. 
However, the equivalent channel is coarse-grained in nature, which limits the SKR. Inspired by the multiple antenna-based and OFDM-based key generation, it is reasonable to envisage generating keys from each RIS-reflected subchannel, which is fine-grained and can provide more diversity. Surprisingly, such research effort is missing.

Joint design of beamforming and RIS will provide more channel diversity for key generation.
Classical key generation systems used multiple-input and multiple-output (MIMO) techniques to improve the SKR \cite{intro20}, which is achieved by beamforming techniques that combine received signals  from  multiple antennas to improve the SNR. Similar to beamforming techniques in MIMO systems, a RIS can achieve passive beamforming to enhance the SNR at the receiver, where a RIS combines the signals from reflecting elements and automatically modifies the reflection coefficients. In \cite{intro10}, the authors proposed to optimize the phase shift matrix to increase the SKR. An optimal selection strategy of RIS elements was designed to maximize SKR~\cite{intro11}. A general analytical expression of the SKR was derived and optimized in multiple-user systems in~\cite{intro12}. However, these works only involved single-antenna BSs hence they did not consider jointly design of the transmit beamforming of the BS and the passive beamforming of the RIS. 

In this paper, we  jointly optimize the precoding matrix at the multiple-antenna base station (BS) and the phase-shift matrix at the RIS and use fine-grained channel feature for key generation. Our main technical contributions are as follows:
\begin{itemize}
    \item We design a channel probing protocol for RIS-assisted key generation systems to fully extract randomness from the direct and cascaded channels. Since the RIS is not capable of estimating channels, our protocol decomposes the channel into the cascaded channel which can be  measured by transceivers.
    
    \item We derive the analytical expression of the SKR in a RIS-assisted multi-antenna system. We optimize the precoding and phase shift matrices to improve the SKR. In order to tackle the coupling problem of two matrix variables, we optimize an equivalent matrix variable that is a combination of the precoding and phase-shift matrices. Furthermore, we introduce a water-filling algorithm based on Karush-Kuhn-Tucker (KKT) conditions to find the optimal equivalent matrix variable.
    
    \item We propose a practical algorithm to decouple the optimal equivalent matrix variable into precoding and phase shift matrices. We first obtain the optimal phase shift matrix and then resort to the Grassmann manifold optimization method to find the optimal precoding matrix. 
	
	\item We validate the analytical expressions of SKR by Monte Carlo simulations. Taking into account of the spatial correlation coefficients, the transmit power, the number of  elements and the Rician factor, we demonstrate that our algorithm achieves a higher SKR than the existing algorithms.
\end{itemize}

The remaining part of this paper is organized as follows. In Section II, we present the system model of the RIS-assisted key generation. In Section III, we propose a RIS-assisted channel probing algorithm. Section IV studies the design of the precoding and phase shift matrices to maximize the SKR. In Section V, numerical results are presented. In Section VI, conclusions are drawn. 

\emph{Notations:} Lower-case letters $(a,b,\dots)$, boldface lower-case letters $(\mathbf{a},\mathbf{b},\dots)$ and boldface upper-case letters $(\mathbf{A},\mathbf{B},\dots)$ denote scalars, vectors and matrices, respectively. Calligraphic letters $(\mathcal{A},\mathcal{B},\dots)$ denote sets. $\frac{\partial f}{\partial x}$ denotes the partial derivative of $f$ with respect to $x$. $\mod(\cdot)$ is the modulus operator and $\lfloor\cdot\rfloor$ is the floor function. $|\cdot|$ and $\Re\{\cdot\}$ denote the magnitude and real part of a complex number, respectively. $\text{diag}(\cdot)$ forms a diagonal matrix out of its vector argument. $\text{vec}(\cdot)$ is the vectorization of a matrix argument. $(\cdot)^T$, $(\cdot)^H$, $(\cdot)^{-1}$ and $(\cdot)^*$ denote the transpose, conjugate transpose, inverse and conjugate, respectively. $\mathbb{C}^{m\times n}$ is the complex space of a $m\times n$ matrix. $\text{Tr}(\cdot)$ is the trace operator. $\mathbf{I}_N$ denotes the $N\times N$ identity matrix. $\boldsymbol{0}_N$ and $\boldsymbol{1}_N$ are the zero and one matrices of $N\times 1$ dimension, respectively. $\mathbf{A}_{i,j} = \mathbf{A}((i-1)m+1:im,(j-1)n+1:jn)$ is the submatrix of $\mathbf{A}$, with row indices spanning from $(i-1)m+1$ to $im$ and column indices spanning from $(j-1)n+1$ to $jn$. $[\mathbf{A}]_{m,n}$ denotes the $(m, n)$-th element of matrix $\mathbf{A}$.  $||\cdot||_2$ and $||\cdot||_F$ denote the Euclidean norm and Frobenius norm, respectively. $\mathcal{CN}(\mu,\sigma^2)$ denotes the circularly symmetric complex Gaussian  distribution with mean $\mu$ and variance $\sigma^2$. $\mathbb{E}\{\cdot\}$ denotes the statistical expectation, and $\otimes$ is the Kronecker product. $\mathbb{H}$ denotes the differential entropy.

\section{System Model}
\subsection{Overview}
\begin{figure}[t!]
  \centering
  \includegraphics[width=3.4in]{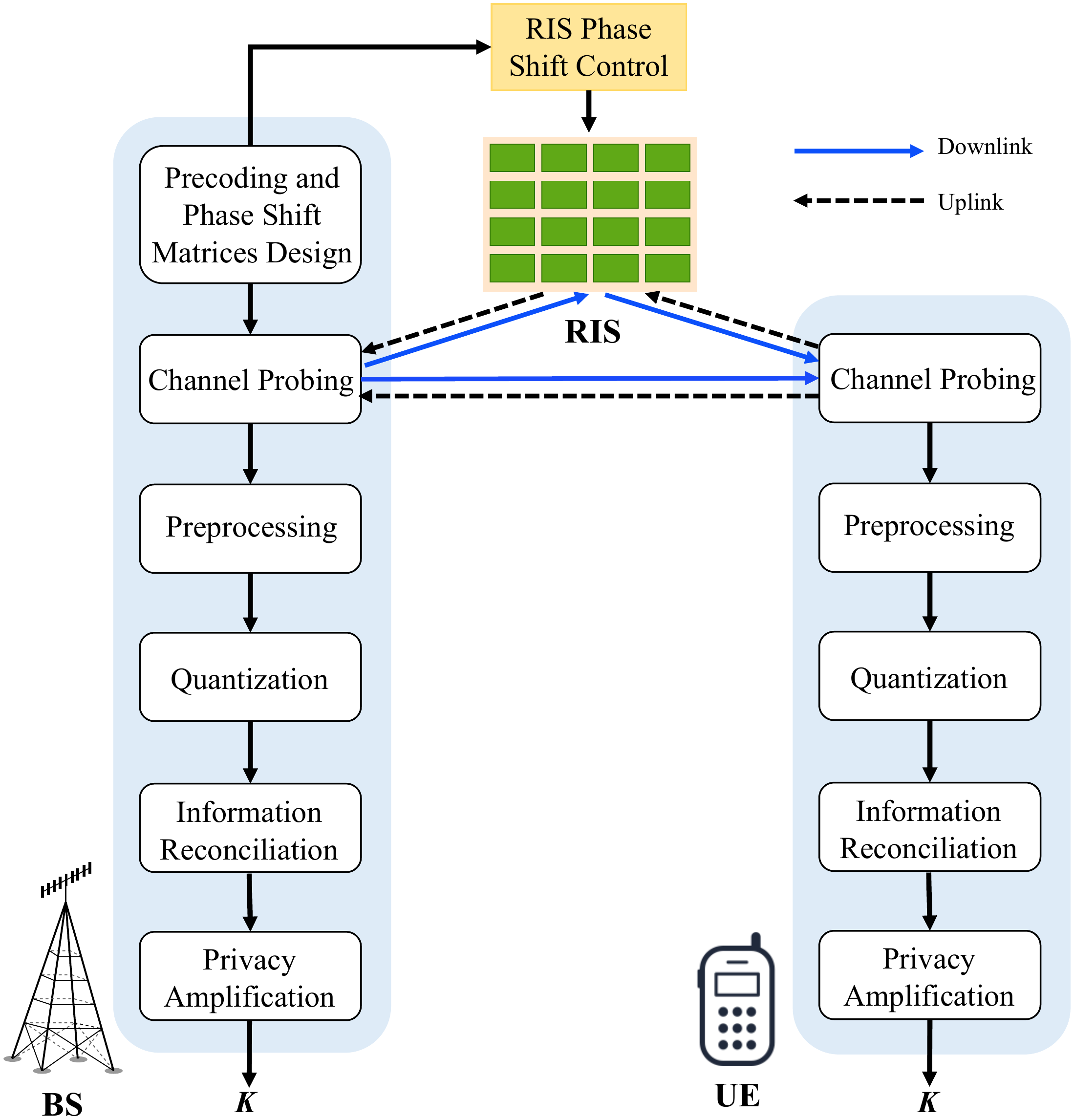}
  \caption{System model.}
  \label{fig.1}
\end{figure}

This paper considers a RIS-assisted key generation system which consists of a BS (Alice),  a user equipment (UE) (Bob), and a RIS, as shown in Fig.~\ref{fig.1}. 
The BS is equipped with multiple antennas, while the UE is equipped with a single antenna. We consider the BS that is linked to a RIS. The BS designs an algorithm for controlling the phase shift matrix of the RIS and its precoding matrix to assist the key generation process, which will be elaborated in Section~\ref{sec:design}.
Key generation protocol comprises five stages, namely channel probing, preprocessing, quantization, information reconciliation and privacy amplification. 
During channel probing, the UE and BS operate in the time division duplex (TDD) mode; they transmit pilots to each other in turn and measure the channels between them. These transmissions will be reflected by the RIS, which can bring more randomness. This paper focuses on the design of channel probing, which will be explained in Section~\ref{sec:channel_probing}.

Since the spatial correlation will cause autocorrelation between measurements, preprocessing methods are introduced to eliminate the autocorrelation between raw channel measurements. Furthermore, they convert the channel measurements into binary sequences by quantization algorithms. 
These two steps will be introduced in Section~\ref{sec:protocol}.

Due to noise, hardware mismatch, etc., there exist disagreements between quantized sequences, which can be corrected during information reconciliation. Finally, the privacy amplification algorithms are used to wipe off possible information leakage in the previous stages. 
The BS and UE agree on a unique secret key, $\bm{K}$. Information reconciliation and privacy amplification are not studied in this paper and are shown in Fig.~\ref{fig.1} for completeness. Interested readers please refer to~\cite{zhang2020new}. 

\subsection{Channel Model}
\subsubsection{Device Configuration} 
\begin{figure}[t!]
  \centering
  \includegraphics[width=2.4in]{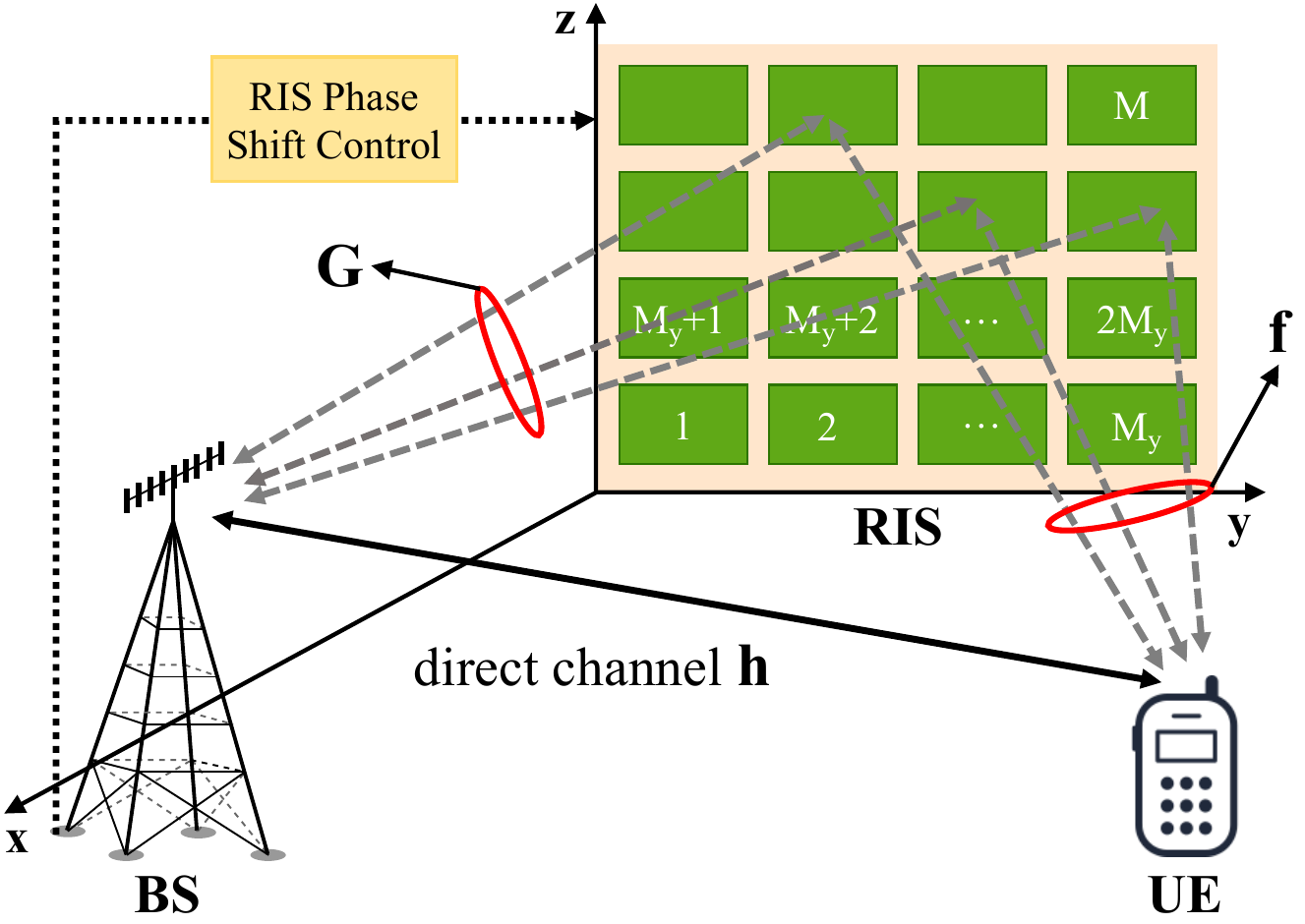}
  \caption{Channel model.}
  \label{Channel model}
\end{figure}

We consider a three-dimensional Cartesian coordinate system, where the RIS is deployed parallel to $y-z$ plane, as shown in Fig. \ref{Channel model}. The RIS is modeled as a uniform planar array that has $M = M_y\times M_z$ reflecting elements with $M_y$ elements per row and $ M_z$ elements per column. The area of the RIS is $MA$, where $d_r$ is the side length of an element and $A = d_r \times d_r$ is the area of an element. The location of the first reflecting element is $\mathbf{u}_1$ and the location of the other elements are denoted as $\mathbf{u}_m = \mathbf{u}_1 + d_r[0, y_m, z_m]^T$, $m = 2,\dots,M$, where $y_m=\mod(m-1, M_y)$ and $z_m=\left \lfloor \frac{m-1}{M_y} \right \rfloor$  are the horizontal and vertical indices, respectively.

When a wave impinges on the RIS from the azimuth angle, $\theta$, and the elevation angle, $\varphi$, the array response vector of the RIS is given by
\begin{align}
    \mathbf{a}(\theta,\varphi)=\mathbf{a}_{z}(\varphi)\otimes\ \mathbf{a}_{y}(\theta,\varphi), 
\end{align}
where $\mathbf{a}_{y}(\theta,\varphi)=[1,\dots,e^{j2\pi(M_z-1)d_r\cos{\varphi}\sin{\theta}/\lambda}]^T$, $\mathbf{a}_{z}(\varphi)=[1,\dots,e^{j2\pi(M_y-1)d_r\sin{\varphi}/\lambda}]^T$ and $\lambda$ is the wavelength.

Each  element of the RIS can control the wave impinging on it. We denote the reflection coefficients of the elements as
\begin{align}
	\mathbf{v}=[\phi_1,\dots,\phi_M]^T,
\end{align}
where $\phi_m$ is the reflection coefficient of the $m$-th element. Specially, $\phi_m=e^{j\omega_m}$, where $\omega_m$ is its phase shift which is generated from uniform quantization of $[0,2\pi)$. The set representing all possible configurations of phase shifts is $\mathcal{K}=\{0,\frac{2\pi}{K_q},\dots,\frac{2\pi(K_q-1)}{K_q}\}$, where $K_q=2^{N_q}$ is the quantization level and $N_q$ is the number of controlling bits.

The BS is equipped with a uniform linear array (ULA) consisting of $N$ antennas and located on the $x$-axis with $d_a$ antenna spacing. When a wave impinges on the ULA from an azimuth angle, $\psi$, the array response vector is given by
\begin{equation}
  \begin{aligned}
\mathbf{b}(\psi)=[1,\dots,e^{j2\pi(N-1)d_a\sin{\psi}/\lambda}]^T.
  \end{aligned}
\end{equation}
The BS applies a precoding matrix, $\mathbf{P}^T\in\mathbb{C}^{N_s\times N}$, for transmission or reception, where $N_s$ is the length of the packet. 


\subsubsection{Individual Channel}
There are three individual channels, including BS-RIS, UE-RIS, and UE-BS channels. The small-scale fading for all channels is assumed to be Rician fading, which is described as follows.

The BS-RIS channel can be modeled as
\begin{equation}
\begin{aligned}
   \mathbf{G} &=\sqrt{\frac{K_{ar}\beta_{ar}}{1+K_{ar}}}\mathbf{G}^{LoS}+\sqrt{\frac{\beta_{ar}}{1+K_{ar}}}\mathbf{G}^{NLoS},
\end{aligned}
\end{equation}
where $\mathbf{G}^{LoS}=\mathbf{a}(\theta_{ar},\varphi_{ar})\mathbf{b}^H(\psi_{ar})$ is the LoS component, $\mathbf{G}^{NLoS}=\sum_l^{L_{ar}}\frac{c_{ar}^l}{\sqrt{L_{ar}}}\mathbf{a}(\theta_{ar}^{l},\varphi_{ar}^l)\mathbf{b}^H(\psi_{ar}^{l})$ is the NLoS component, $K_{ar}$ is the Rician factor and $\beta_{ar}$ is the distance-dependent path-loss effect. 
\begin{itemize}
	\item For the LoS component from the BS to the RIS, $\psi_{ar}\in[0,2\pi)$ denotes the angle of departure (AoD). $\theta_{ar}\in[0,2\pi)$ and $\varphi_{ar}\in[-\pi/2,\pi/2)$ denote the azimuth and elevation angles of arrival (AoA), respectively.
	\item For the NLoS component from the BS to the RIS, $L_{ar}$ is the number of paths and $c_{ar}^l$ denotes the corresponding complex gain associated with the $l$-th path. $\psi_{ar}^l$ denotes the AoD of the $l$-th path. $\theta_{ar}^{l}$ and $\varphi_{ar}^{l}$ denote the azimuth and elevation AoA of the $l$-th path, respectively. The complex gain $c^l_{ar}$ is identically and independently distributed (i.i.d.) with zero mean and variance $A\mu_1$, where $\mu_1$ is the average intensity attenuation. 
\end{itemize}
 
The UE-RIS channel is modeled as
\begin{align}
    \mathbf{f}\!=\! \underbrace{\sqrt{\frac{K_{br}\beta_{br}}{1\!+\!K_{br}}}\mathbf{a}_{r}(\theta_{br},\varphi_{br})}_{LoS}
		\!+\!\underbrace{\sqrt{\frac{\beta_{br}}{1\!+\!K_{br}}}\!\!\sum\limits^{L_{br}}_{l=1}\!\frac{c_{br}^{l}}{\sqrt{L_{br}}}\mathbf{a}_{br}(\theta^{l}_{br},\varphi^{l}_{br})}_{NLoS},
\end{align}
where $K_{br}$ is the Rician factor and $\beta_{br}$ is the path-loss effect. 
\begin{itemize}
	\item For the LoS component from the UE to the RIS, $\theta_{br}$ and $\varphi_{br}$ denote the azimuth and elevation AoA, respectively. 
	\item For the NLoS component, $\theta_{br}^{l}$ and $\varphi_{br}^{l}$ denote the azimuth and elevation AoA of the $l$-th path, respectively. $L_{br}$ is the number of paths and $c_{br}^{l}$ denotes the complex gain associated with the $l$-th path.
\end{itemize}

Similarly, we model the UE-BS channel, also termed direct channel in this paper, as
\begin{align}
    \mathbf{h}= \underbrace{\sqrt{\frac{K_{ba}\beta_{ba}}{1+K_{ba}}}\mathbf{b}(\psi_{ba})}_{LoS}+\underbrace{\sqrt{\frac{\beta_{ba}}{1+K_{ba}}}\sum\limits_{l=1}^{L_{ba}}\frac{c_{ba}^{l}}{\sqrt{L_{ba}}}\mathbf{b}(\psi^{l}_{ba})}_{NLoS},
\end{align}
where $K_{ba}$ is the Rician factor and $\beta_{ba}$ is the path-loss effect. 
\begin{itemize}
	\item For the LoS component from the UE to the BS, $\psi_{ba}$ denotes the azimuth AoA. 
	\item For the NLoS component, $\varphi_{ba}^{l}$ denotes the azimuth AoA of the $l$-th path. $L_{ba}$ is the number of paths and $c_{ba}^{l}$ denotes the complex gain associated with the $l$-th path.
\end{itemize}

\subsubsection{Cascaded Channel}
When the RIS configures the phase shift vector, $\mathbf{v}$, the BS and UE can observe the combined version of individual channels, namely $\mathbf{h}$, $\mathbf{f}$ and $\mathbf{G}$. Given $\mathbf{v}$, we define the equivalent channel, $\mathbf{h}_{e}(\mathbf{v})\in\mathbb{C}^{N\times 1}$, as
\begin{align}\label{eq6}
    \mathbf{h}_{e}(\mathbf{v}) = \mathbf{h} + \mathbf{G}^T\text{diag}(\mathbf{v})\mathbf{f},    
\end{align}
where $\mathbf{h}\in\mathbb{C}^{N\times 1}$, $\mathbf{G}^T=[\mathbf{g}_1,\dots,\mathbf{g}_M]\in \mathbb{C}^{N\times M}$, and $\mathbf{f}=[f_{1},\dots,f_{M}]^T\in \mathbb{C}^{M\times 1}$. Here, $\mathbf{g}_m\in\mathbb{C}^{N\times 1}$, $m=1,\dots,M$, is the channel from the $m$-th  element to the BS. The equivalent channel is composed of $\mathbf{h}$ and $\mathbf{G}^T\text{diag}(\mathbf{v})\mathbf{f}$, where $\mathbf{h}$ is the UE-BS channel that does not involve the RIS and $\mathbf{G}^T\text{diag}(\mathbf{v})\mathbf{f}$ is the channel that can be adjusted by the RIS.
The BS and UE can directly estimate the equivalent channel and convert the measurements to secret keys, which is commonly adopted by previous works~\cite{intro10,intro11,intro12}. However, the equivalent channel is coarse-grained, which fundamentally limits the SKR. 

The equivalent channel can be decomposed into the UE-BS channel, $\mathbf{h}$, and a set of channels associated with each element, $\mathbf{h}_m$. The equivalent channel in \eqref{eq6} can be rewritten as
\begin{align}
    \mathbf{h}_{e}(\mathbf{v}) = \mathbf{h} + \mathbf{G}^T\text{diag}(\mathbf{f})\mathbf{v}.  
\end{align}
$\mathbf{G}^T\text{diag}(\mathbf{f})=[\mathbf{h}_{1}\dots,\mathbf{h}_{M}]$ is the channel associated with  elements, where $\mathbf{h}_{m}=\mathbf{g}_{m}f_{m}$ is the channel from the UE to the $m$-th element of the RIS and then to the BS. Therefore, we propose to exploit the randomness from the UE-BS channel and the channels associated with each  element, which is finer-grained and can improve the SKR. 

Although  $\mathbf{h}$ and $\mathbf{h}_m$ provide more channel coefficients for extracting secret keys, there is spatial correlation between  $\mathbf{h}_m$ and $\mathbf{h}_n$, $m\neq n$ or between $\mathbf{h}$ and $\mathbf{h}_m$. Since the spatial correlation influences the actual secret keys, we cannot simply estimate these channels, individually quantize the measurements from each channel and then convert it to the secret keys. To analyze the secret keys influenced by the spatial correlation, we stack $\mathbf{h}$ and $\mathbf{h}_{m}$, $m = 1,\dots,M$, into $\mathbf{h}_{r}=[\mathbf{h}^T,\mathbf{h}_{1}^T\dots,\mathbf{h}_{M}^T]^T$. We define $\mathbf{h}_{r}\in\mathbb{C}^{D\times 1}$ as the BS-RIS-UE cascaded channel, which is given by
    \begin{align}\label{eq3}
       \mathbf{h}_{r}=[\mathbf{h}^T,\mathbf{h}_{1}^T\dots,\mathbf{h}_{M}^T]^T=\text{vec}([\mathbf{h}~\mathbf{G}^T\text{diag}(\mathbf{f})]),
    \end{align}
where $D=N(M+1)$ is the dimension of the cascaded channel. We define the $m$-th entry of $\mathbf{h}_{r}$ as the $m$-th subchannel. $\mathbf{h}_{r}$ is composed of $D$ subchannels.

Since the RIS is passive, we cannot directly estimate $\mathbf{G}$ and $\mathbf{f}$ at the RIS and construct the cascaded channel. In order to measure the cascaded channel, we decompose the equivalent channel as
\begin{align}\label{equation3}
       \mathbf{h}_{e}(\mathbf{v}) & \overset{(a)}{=}\begin{bmatrix}\mathbf{h}~ \mathbf{G}^T\text{diag}(\mathbf{f})\end{bmatrix}
       \begin{bmatrix}
         1 \\
        \mathbf{v}
       \end{bmatrix}      
    \overset{(b)}{=}(\bar{\mathbf{v}}^T\otimes\mathbf{I}_{N})\mathbf{h}_{r},
\end{align}
where $(a)$ holds due to $\text{diag}(\mathbf{a})\mathbf{b}=\text{diag}(\mathbf{b})\mathbf{a}$, $(b)$ holds due to $\text{vec}(\mathbf{A}\mathbf{B}\mathbf{C})=(\mathbf{C}^T\otimes \mathbf{A})\text{vec}(\mathbf{B})$ \cite{bib4} and $\bar{\mathbf{v}}=[1,\mathbf{v}^T]^T$ is the augmented phase shift vector. In Section \ref{sec:channel_probing}, we will propose a channel probing protocol in which the UE and BS design $\bar{\mathbf{v}}$ in \eqref{equation3} to measure the cascaded channel and extract the randomness from it.

\subsection{Correlation Modeling}
The side length of an element and the antenna spacing affect the spatial correlation between subchannels of the cascaded channel. In order to model the spatial correlation between subchannels, we analyze the correlation matrices between transmit antennas and elements.

The NLoS channel observed from RIS is spatially correlated. According to \cite{bib1}, the correlation matrix of the elements, $\mathbf{R}_r$, in isotropic scattering environments is modelled as
\begin{equation}
    [\mathbf{R}_r]_{n,m}=\gamma\frac{\sin(\frac{2\pi}{\lambda}||\mathbf{u}_n-\mathbf{u}_m||_2)}{\frac{2\pi}{\lambda}||\mathbf{u}_n-\mathbf{u}_m||_2}~n,m=1,\dots,M,
\end{equation} 
where $[\mathbf{R}_r]_{n,m}$ is the entry in the $n$-th row and the $m$-th column of $\mathbf{R}_r$ and $\gamma = \frac{1}{A\mu_1}$ is a normalizing factor.

There is correlation between  transmit antennas. According to \cite{bib3}, the correlation coefficient between the $i$-th and $j$-th transmit antennas is given by $[\mathbf{R}_a]_{i,j}=r^{|i-j|}$, where $r$ is the correlation coefficient and $\mathbf{R}_a$ is the covariance matrix. Therefore, $\mathbf{h}\sim\mathcal{CN}(\bm{0},\mathbf{R}_a^{1/2})$, $\mathbf{f}\sim\mathcal{CN}(\bm{0},\mathbf{R}_r^{1/2})$ and $\mathbf{G}=\mathbf{R}_r^{1/2}\mathbf{H}\mathbf{R}_{a}^{1/2}$, where $\mathbf{H}\in\mathbb{C}^{M\times N}$ is the random muti-path CSI matrix with identically independent distributed (i.i.d.) entries.

According to \eqref{eq3}, the spatial correlation matrix of the BS-RIS-UE cascaded channel, $\mathbf{h}_{r}$, can be derived from $\mathbf{R}_{h}=\mathbb{E}\{\mathbf{h}_{r}\mathbf{h}_{r}^H\}$, $\mathbf{R}_h\in\mathbb{C}^{D\times D}$. We assume the BS has the knowledge of the spatial covariance matrix. Let $\mathbf{R}_h=\mathbf{U}_h\mathbf{\Lambda}_h\mathbf{U}_h^H$ denote the eigenvalue decomposition of $\mathbf{R}_h$. Specially, $\mathbf{\Lambda}_h$ is the diagonal matrix whose diagonal entries are the eigenvalues of $D$ subchannels and $\mathbf{U}_h$ is a unitary matrix whose columns are eigenvectors corresponding to eigenvalues. We define $\mathbf{\Lambda}_h=\text{diag}(\mathbf{p}_h)$ with $\mathbf{p}_h = [p_{h,1},p_{h,2},\dots,p_{h,D}]$, where the elements in $\mathbf{p}_h$ are sorted in the descending order.

\section{RIS Assisted Channel Probing}\label{sec:channel_probing}
In this section, a channel probing protocol is designed to measure the cascaded channel of \eqref{eq3} from which the secret keys are extracted. The main difficulty is that a RIS can neither send nor receive pilots since it does not have any radio resources. Also, the dimension of the cascaded channel is $D$, so the BS and UE cannot measure it by only conducting one round of channel probing. However, the UE and BS can obtain partial information on the cascaded channel from \eqref{equation3} when they estimate the equivalent channel, so they can collect multiple measurements of the equivalent channel to recover the cascaded channel. The phase shift vector and precoding matrix are appropriately configured in each round. Furthermore, we derive the SKR of the measurements of the cascaded channel.  

\subsection{Channel Probing}
\begin{figure}[t]
  \centering
  \includegraphics[width=3.4in]{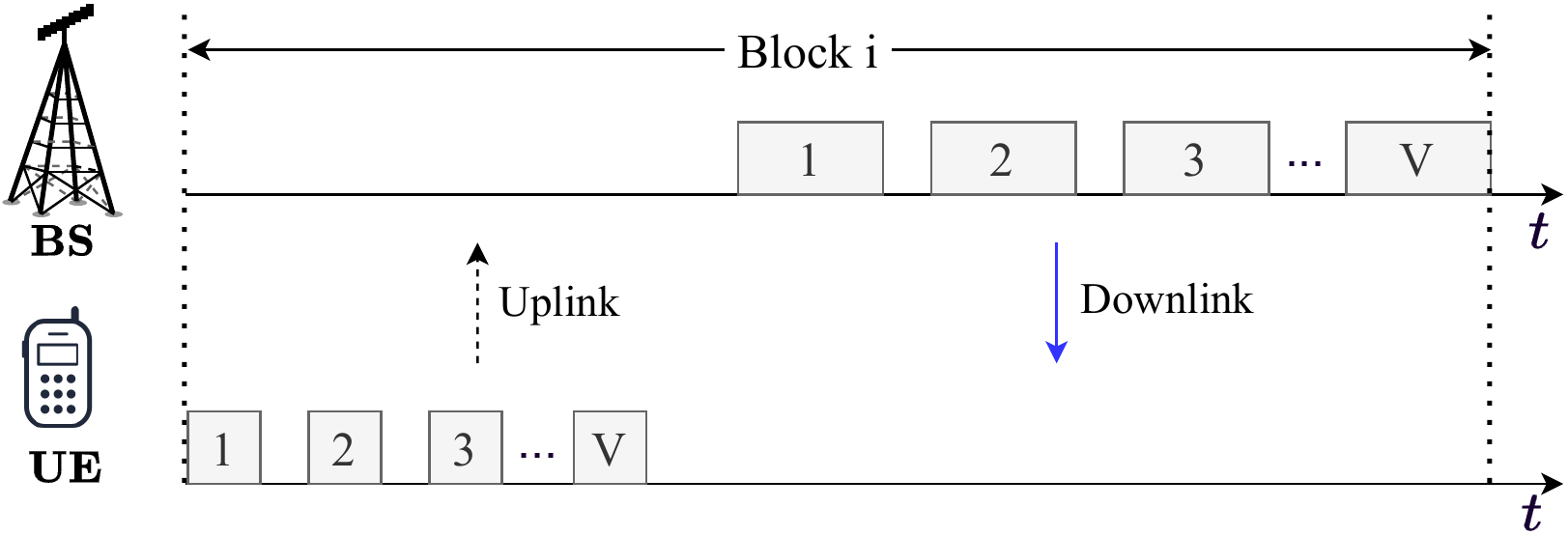}
  \caption{Channel probing in a fading block.  }
  \label{fig.2}
\end{figure}

As shown in Fig. \ref{fig.2}, the channel probing consists of uplink and downlink phases. In the uplink (downlink) phase, the UE (BS) transmits $V$ packets to BS (UE) in fading block $i$. 
In order to recover the cascaded channel, the BS configures a combination of $V$ phase shift vectors for $V$ packets in the uplink or downlink phases, where the combination in the uplink phase is the same as the combination in the downlink phase.

In the uplink phase, each  packet is $\mathbf{s}\in\mathbb{C}^{Q\times 1}$ of length $Q$. After receiving the uplink packet, the BS measures the channel from $N$ antennas and applies a $N_s\times N$ precoding matrix, $\mathbf{P}^T$, to the measurements. Since the BS is equipped with $N$ antennas, we set $Q = 1$ to let the BS observe $N$ channel coefficients from an uplink packet. In the downlink phase, each downlink packet, $\mathbf{P}\mathbf{S}_{d}^H$, is derived from multiplying the downlink pilot matrix, $\mathbf{S}_d\in\mathbb{C}^{N_s \times N_s}$, by the $\mathbf{P}$. In order to let the UE observe the $N$ channel coefficients, we set $N_s$ as $N$. 

\subsubsection{Uplink Channel Probing}
In the $t$-th uplink packet, the UE transmits an uplink packet to the BS when the RIS configures the phase shift vector $\mathbf{v}(t)$ to reflect it. The received signal at the BS (Alice) is given by
  \begin{align} 
    \mathbf{y}_a(t) =\mathbf{h}_e(\mathbf{v}(t))\mathbf{s}^H+\mathbf{n}_a(t), 
  \end{align}
where $\mathbf{y}_a(t)\in\mathbb{C}^{N\times 1}$ and $\mathbf{n}_a(t)\in\mathbb{C}^{N\times 1}$ is the complex Gaussian noise.

The BS then applies $\mathbf{P}^T$ to transform $\mathbf{y}_a(t)$ as
  \begin{align} \label{eq13}
    \widetilde{\mathbf{y}}_{a}(t) = \mathbf{P}^T\mathbf{y}_{a}(t)
    = \mathbf{P}^T\mathbf{h}_e(\mathbf{v}(t))\mathbf{s}^H+\mathbf{P}^T\mathbf{n}_a(t),
  \end{align}
where $\widetilde{\mathbf{y}}_a(t)\in\mathbb{C}^{N \times 1}$.

By the least square (LS) channel estimation, the BS measures the equivalent channel as
\begin{align}
    \widehat{\mathbf{z}}_{a}(t) =\widetilde{\mathbf{y}}_{a}(t)\mathbf{s}(\mathbf{s}^H\mathbf{s})^{-1}, 
\end{align}
where $\widehat{\mathbf{z}}_{a}(t)\in\mathbb{C}^{N \times 1}$. According to \eqref{equation3}, the measurement obtained from the $t$-th packet can be expanded as
\begin{align}\label{eq14}
  \widehat{\mathbf{z}}_{a}(t) & = \mathbf{P}^T\mathbf{h}_e(\mathbf{v}(t))\mathbf{s}^H\mathbf{s}(\mathbf{s}^H\mathbf{s})^{-1}+\mathbf{P}^T\mathbf{n}_a(t)\mathbf{s}(\mathbf{s}^H\mathbf{s})^{-1} \nonumber \\
  & = \mathbf{P}^T(\bar{\mathbf{v}}^T(t)\otimes\mathbf{I}_{N})\mathbf{h}_{r} +\frac{1}{P_b}\mathbf{P}^T\mathbf{n}_a(t)\mathbf{s} \nonumber\\
  & = (\bar{\mathbf{v}}^T(t)\otimes\mathbf{P}^T)\mathbf{h}_{r} +\frac{1}{\sqrt{P_b}} \widehat{\mathbf{n}}_{a}(t),
  \end{align}
where $\widehat{\mathbf{n}}_{a}(t)=\mathbf{P}^T\mathbf{N}(t)\widetilde{\mathbf{s}}$ and $P_b$ is UE's transmit power. $\widetilde{\mathbf{s}}=\frac{1}{\sqrt{P_b}}\mathbf{s}$ is the normalized uplink packet which has unit transmit power, i.e., $\mathbb{E}\{\widetilde{\mathbf{s}}\widetilde{\mathbf{s}}^H\}=\mathbf{I}$. Since the $\mathbf{P}$ is unitary, $\widehat{\mathbf{n}}_{a}(t)$ has covariance matrix, $\sigma_a^2\mathbf{I}_{N_s}$.

We define $\bar{\mathbf{\Phi}}=[\bar{\mathbf{v}}(1),\dots,\bar{\mathbf{v}}(V)]\in\mathbb{C}^{(M+1)\times V}$ as the phase shift matrix to model the configuration of the phase shift vector over $V$ packets. After collecting the $V$ measurements in \eqref{eq14} and stacking them into a vector, the BS obtains the measurement of the cascaded channel, which is given by
  \begin{align}\label{eq15}
    \mathbf{z}_{a} & =[\widehat{\mathbf{z}}_{a}^T(1),\dots,\widehat{\mathbf{z}}_{a}^T(V)]^T \nonumber\\
    &=(\bar{\mathbf{\Phi}}^T\otimes\mathbf{P}^T)\mathbf{h}_{r} +\frac{1}{\sqrt{P_b}}\boldsymbol{\eta}_{a},
  \end{align}
where $\mathbf{z}_{a}\in\mathbb{C}^{VN\times1}$, and $\boldsymbol{\eta}_{a}=[\widehat{\mathbf{n}}_{a}^T(1),\dots,\widehat{\mathbf{n}}_{a}^T(V)]^T$ with covariance matrix, $\sigma_a^2\mathbf{I}_{VN}$. The cascaded channel consists of $D$ subchannels. Therefore, $NV\geq D$ should be satisfied to meet the requirement of the channel dimension. With $D=N(M+1)$, we have $V\geq (M+1)$.

\subsubsection{Downlink Channel Probing}
In the $t$-th downlink packet, the BS applies $\mathbf{P}$ to send a downlink packet, $\mathbf{P}\mathbf{S}_{d}^H$, to the UE. With the assumption of equal power allocation, we have $\mathbf{S}_{d}^H\mathbf{S}_{d} = NP_a\mathbf{I}_{N}$, where $P_a$ is BS's transmit power. The RIS configures $\mathbf{v}(t)$ to reflect the downlink packet and then the UE (Bob) gets the received signal as
\begin{align}
    \mathbf{y}_{b}(t)&=\mathbf{h}_{e}^T(\mathbf{v}(t))\mathbf{P}\mathbf{S}_{d}^H+\mathbf{n}_{b}(t),                           
\end{align}
where $\mathbf{y}_{b}(t)\in \mathbb{C}^{1\times N}$ and $\mathbf{n}_{b}(t)\in\mathbb{C}^{1\times N}$ is the noise vector.

By the LS estimation, the UE measures the equivalent channel as
\begin{align}
    \widehat{\mathbf{z}}_{b}(t)&= \mathbf{y}_{b}(t)\mathbf{S}_{d}(\mathbf{S}_{d}^H\mathbf{S}_{d})^{-1},
\end{align}
where $\widehat{\mathbf{z}}_{b}(t)\in\mathbb{C}^{1\times N}$.

Based on \eqref{equation3}, the measurement can be expanded as
  \begin{align}
     \widehat{\mathbf{z}}_{b}(t)&= \mathbf{h}_{e}^T(\mathbf{v}(t))\mathbf{P}+\mathbf{n}_{b}(t)\mathbf{S}_{d}(\mathbf{S}_{d}^H\mathbf{S}_{d})^{-1}          \nonumber\\
     &=\mathbf{h}_{r}^T(\bar{\mathbf{v}}(t)\otimes\mathbf{I}_{N})\mathbf{P}+\mathbf{n}_{b}(t)\mathbf{S}_{d}(\mathbf{S}_{d}^H\mathbf{S}_{d})^{-1}\nonumber\\
     & =\mathbf{h}_{r}^T(\bar{\mathbf{v}}(t)\otimes\mathbf{P})+\frac{1}{\sqrt{NP_a}}\widehat{\mathbf{n}}_{b}(t),
  \end{align}
where $\widehat{\mathbf{n}}_{b}(t)\in\mathbb{C}^{1\times N}$, $\widehat{\mathbf{n}}_{b}(t)=\mathbf{n}_{b}(t)\widetilde{\mathbf{S}}_d$ and $\widetilde{\mathbf{S}}_d$ is the normalization of $\mathbf{S}_d$ with $\mathbf{S}_d=\frac{1}{\sqrt{NP_a}}\widetilde{\mathbf{S}}_d$.

The UE computes the transpose of the above equation and replaces $\widehat{\mathbf{z}}_{b}(t)$ with $\widehat{\mathbf{z}}_{b}^T(t)$, which is given by
\begin{align}\label{eq18}
    \bar{\mathbf{z}}_{b}(t)=\widehat{\mathbf{z}}_{b}^T(t) =(\bar{\mathbf{v}}^T(t)\otimes\mathbf{P}^T)\mathbf{h}_{r} +\widehat{\mathbf{n}}^T_{b}(t).
\end{align}

After collecting $V$ measurements in \eqref{eq18} and stacking them into a vector, the UE obtains
\begin{align}\label{eq20}
  \mathbf{z}_{b}  &=[\bar{\mathbf{z}}_{b}(1)^T,\dots,\bar{\mathbf{z}}_{b}^T(V)]^T\nonumber\\
  &=(\bar{\mathbf{\Phi}}^T\otimes\mathbf{P}^T)\mathbf{h}_{r} + \frac{1}{\sqrt{NP_a}}\boldsymbol{\eta}_{b},
\end{align}
where $\mathbf{z}_{b}\in\mathbb{C}^{VN\times1}$ and $\boldsymbol{\eta}_{b}=[\mathbf{n}^T_{b}(1),\dots,\mathbf{n}^T_{b}(V)]^T$ with $\mathbb{E}\{\boldsymbol{\eta}_{b}\boldsymbol{\eta}_{b}^H\}=\sigma_b^2\mathbf{I}_{VN}$. 

\subsection{Secret Key Rate}
After channel probing, the BS and UE distill secret keys from their measurements, $\mathbf{z}_a$, and $\mathbf{z}_b$, respectively. We assume eavesdroppers are more than half-wavelength away from the legitimate terminals, which means Eve's measurements are independent of BS's or UE's measurements. Therefore, the SKR is $I(\mathbf{z}_a;\mathbf{z}_b)$, i.e., the mutual information of the measurements.

In order to calculate the $I(\mathbf{z}_a;\mathbf{z}_b)$, we construct the covariance matrix of the measurements, $\mathbf{z}_{a}$ and $\mathbf{z}_{b}$. Let $\mathbf{R}_{\mathbf{z}_{a}}=\mathbb{E}\{\mathbf{z}_{a}\mathbf{z}_{a}^H\}$ and $\mathbf{R}_{\mathbf{z}_{b}}=\mathbb{E}\{\mathbf{z}_{b}\mathbf{z}_{b}^H\}$ denote the covariance matrices of BS's and UE's measurements, respectively.
Let $\mathbf{R}_{\mathbf{z}_{a}\mathbf{z}_{b}}=\mathbb{E}\{\mathbf{z}_{a}\mathbf{z}_{b}^H\}$ denote the cross-covariance matrix of $\mathbf{z}_{a}$ and $\mathbf{z}_{b}$. According to \eqref{eq15} and \eqref{eq20}, $\mathbf{R}_{\mathbf{z}_{a}}$, $\mathbf{R}_{\mathbf{z}_{b}}$ and $\mathbf{R}_{\mathbf{z}_{a}\mathbf{z}_{b}}$ can be expressed as 
\begin{align}\label{eq3.1}
    \mathbf{R}_{\mathbf{z}_{a}}&= (\bar{\mathbf{\Phi}}\otimes\mathbf{P})^T\mathbf{R}_h(\bar{\mathbf{\Phi}}\otimes\mathbf{P})^*+\frac{\sigma_a^2}{P_b}\mathbf{I}_D,\\
        \mathbf{R}_{\mathbf{z}_{b}}&=(\bar{\mathbf{\Phi}}\otimes\mathbf{P})^T\mathbf{R}_h(\bar{\mathbf{\Phi}}\otimes\mathbf{P})^*+\frac{\sigma_b^2}{NP_a}\mathbf{I}_D,\\
     \mathbf{R}_{\mathbf{z}_{a}\mathbf{z}_{b}}&=\mathbf{R}_{\mathbf{z}_{b}\mathbf{z}_{a}}=(\bar{\mathbf{\Phi}}\otimes\mathbf{P})^T\mathbf{R}_h(\bar{\mathbf{\Phi}}\otimes\mathbf{P})^*.
\end{align}

The full covariance matrix of both measurements is
\begin{equation}\label{eq3.2}
  \mathbf{K}_{\mathbf{z}_{a}\mathbf{z}_{b}}=\mathbb{E}\left\{\begin{bmatrix} \widehat{\mathbf{z}}_a\\\widehat{\mathbf{z}}_b\end{bmatrix}[\widehat{\mathbf{z}}_a^H,\widehat{\mathbf{z}}_b^H]\right\}
  =\begin{bmatrix} \mathbf{R}_{\mathbf{z}_{a}}~\mathbf{R}_{\mathbf{z}_{a}\mathbf{z}_{b}} \\
    \mathbf{R}_{\mathbf{z}_{b}\mathbf{z}_{a}}~\mathbf{R}_{\mathbf{z}_{b}}
  \end{bmatrix}.
\end{equation}

The SKR is given by
\begin{align}
   I(\mathbf{z}_{a};\mathbf{z}_{b}) & =\mathbb{H}(\mathbf{z}_{a})+\mathbb{H}(\mathbf{z}_{b})-\mathbb{H}(\mathbf{z}_{a},\mathbf{z}_{b})    \nonumber\\
     & =\log_2 \left(\frac{|\mathbf{R}_{\mathbf{z}_{a}}||\mathbf{R}_{\mathbf{z}_{b}}|}{|\mathbf{K}_{\mathbf{z}_{a}\mathbf{z}_{b}}|}\right)                                                                   \nonumber\\
      & \overset{(a)}{=}\log_2 \Big(\frac{|\mathbf{R}_{\mathbf{z}_{b}}|}{|\mathbf{R}_{\mathbf{z}_{b}}-\mathbf{R}_{\mathbf{z}_{b}\mathbf{z}_{a}}\mathbf{R}_{\mathbf{z}_{a}}^{-1}\mathbf{R}_{\mathbf{z}_{a}\mathbf{z}_{b}}|}\Big),
\end{align}
where $(a)$ holds due to determinant of the block matrix. Substituting \eqref{eq3.1}-\eqref{eq3.2} into the above equation, we have 
\begin{align}\label{ob}
    &I(\mathbf{z}_{a};\mathbf{z}_{b})=\log_2\left(\frac{|\mathbf{R}_W+\mathbf{\Gamma}_b|}{|\mathbf{R}_W+\mathbf{\Gamma}_b-\mathbf{R}_W(\mathbf{R}_W+\mathbf{\Gamma}_a)^{-1}\mathbf{R}_W|}\right) \nonumber\\
    & =-\log_2\left(|\mathbf{I}-\mathbf{R}_W(\mathbf{R}_W+\mathbf{\Gamma}_a)^{-1}\mathbf{R}_W(\mathbf{R}_W+\mathbf{\Gamma}_b)^{-1}|\right),
\end{align}
where $\mathbf{R}_W=(\bar{\mathbf{\Phi}}\otimes\mathbf{P})^T\mathbf{R}_h(\bar{\mathbf{\Phi}}\otimes\mathbf{P})^*$.

\section{Secret Key Rate Optimization}
The objective is to jointly design the phase shift matrix, $\bar{\mathbf{\Phi}}$, and the precoding matrix, $\mathbf{P}$,  to maximize the SKR when the BS has the information of the channel spatial covariance matrix. We formulate the following optimization problem:
\begin{align}
  \text{(P1)}~\mathop{\max_{\bar{\mathbf{\Phi}},\mathbf{P}}}~&I(\mathbf{z}_{a};\mathbf{z}_{b})     \nonumber\\
      s.t.~&|\phi_{m,t}|=1, \label{c1}\\
           &\bar{\phi}_{m,t}\in\mathcal{K},~2\leq m\leq M+1,1\leq t\leq V,\\
           &\bar{\phi}_{1,t}=1,~1\leq t\leq V,\\\label{c2}
           &\text{rank}(\bar{\boldsymbol{\Phi}})=M+1,\\\label{c3}
           &\mathbf{P}^H\mathbf{P}= \mathbf{I}_N.
\end{align}

The phase shift and precoding matrices, $\bar{\mathbf{\Phi}}$ and $\mathbf{P}$, should meet the following five constraints. 
\begin{itemize}
    \item The RIS is passive and each reflection coefficient cannot modify the amplitude but the phase shift. The RIS should meet the unit-module constraint of $|\phi_{m,t}|=1$.
	\item The $m$-th row and the $t$-th column entry of the $\bar{\mathbf{\Phi}}$ should meet $\bar{\phi}_{m,t}\in\mathcal{K}$, $2\leq m\leq (M+1)$, $1\leq t\leq V$, which means that the RIS can select the reflection coefficient of the $m$-th element from $\mathcal{K}$ in the $t$-th packet.
	\item The first row of the phase shift matrix, $\bar{\mathbf{\Phi}}$, represents the phase shift that the RIS configured for the direct channel. However, the direct channel which does not pass through the RIS is uncontrollable, so the first row of $\bar{\mathbf{\Phi}}$ should meet $\bar{\phi}_{1,t}=1,~1\leq t\leq V$.
	\item The equation $\text{rank}(\bar{\boldsymbol{\Phi}})=M+1$ should satisfy to ensure the BS and UE can measure the cascaded channel. 
	\item With the assumption of equal power allocation, the precoding matrix should meet the constraint of $\mathbf{P}^H\mathbf{P}= \mathbf{I}_N$. 
\end{itemize}

Note that (P1) is difficult to solve because the constraints of \eqref{c1} and \eqref{c2} are non-convex and the Kronecker product structure in the objective function. Therefore, we transform the $\bar{\mathbf{\Phi}}\otimes\mathbf{P}$ into an equivalent matrix variable, $\mathbf{W}\in\mathbb{C}^{D\times (N_sV)}$, and relax some constraints to formulate an approximate optimization problem. We design an algorithm to find the optimal solution, $\mathbf{W}^{\dag}$, of the approximate optimization problem, as presented in Section~\ref{sec:upper_bound}. $\mathbf{W}^{\dag}$ reaches the upper bound of (P1) but it cannot be applied in practice. Therefore, we will design an algorithm to divide $\mathbf{W}^{\dag}$ into $\bar{\mathbf{\Phi}}$ and $\mathbf{P}$, which is done in Section~\ref{sec:design}.

\subsection{Upper Bound of SKR}\label{sec:upper_bound}
The rank constraint of the phase shift matrix and the unit-modulus constraint of the reflection coefficient are non-convex. Therefore, we relax the constraints and transform (P1) into (P2) to find the upper bound of the problem (P1). The (P1) can be reformulated as 
\begin{equation}\label{ob28}
   \begin{aligned}
\text{(P2)}~\mathop{\max_{\mathbf{W}}}~&I(\mathbf{z}_{a};\mathbf{z}_{b})  \\
     s.t.~&\|\mathbf{W}\|_F^2 = (M+1)^2N.
  \end{aligned}
\end{equation}

Next, we will discuss how to find the optimal $\mathbf{W}$. Firstly, to simplify the objection function of (P2), we transform the matrix variable, $\mathbf{W}$, into a set of scalar variables, $p_i$, $i=1,\dots,D$. Secondly, we analyze the concavity of the objective function in terms of $p_i$. Finally, we derive the KKT condition of the optimization problem based on the concavity and design an algorithm to find the optimal $p_i$, which allows us to find the optimal $\mathbf{W}$.

\subsubsection{Converting $\mathbf{W}$ to Scalar Variables $p_i$} Define $\mathbf{R}_W=\mathbf{W}^T\mathbf{R}_h\mathbf{W}^*$. We do the Cholesky factorization of the spatial correlation matrix, $\mathbf{R}_h$, as $\mathbf{R}_h=\mathbf{R}_h^{1/2}(\mathbf{R}_h^{1/2})^H$, where $\mathbf{R}_h^{1/2}=\mathbf{U}_h\mathbf{\Lambda}_h^{1/2}$, $(\mathbf{R}_h^{1/2})^H=\mathbf{\Lambda}_h^{1/2}\mathbf{U}_h^H$ and $\mathbf{R}_W=\mathbf{W}^T\mathbf{R}_h^{1/2}(\mathbf{R}_h^{1/2})^H\mathbf{W}^*=((\mathbf{R}_h^{1/2})^H\mathbf{W}^*)^H((\mathbf{R}_h^{1/2})^H\mathbf{W}^*)$. To simplify the objective function of (P2), we resort to the following singular value decomposition (SVD).
\begin{align}
    &\mathbf{R}_h^{H/2}\mathbf{W}^*=\mathbf{U}\mathbf{\Lambda}\mathbf{V}^H,\label{eq29}\\
    &\mathbf{W}=(\mathbf{R}_h^{-H/2}\mathbf{U}\mathbf{\Lambda}\mathbf{V}^H)^*=(\mathbf{U}_h\mathbf{\Lambda}_h^{-1/2}\mathbf{U}\mathbf{\Lambda}\mathbf{V}^H)^*,\label{eq30}\\
    &\mathbf{R}_W=\mathbf{V}\mathbf{\Lambda}\mathbf{\Lambda}^T\mathbf{V}^H, \label{eq31}
\end{align}
where $\mathbf{U}\in\mathbb{C}^{D\times D}$, $\mathbf{\Lambda}\in\mathbb{C}^{D\times N_sV}$ and $\mathbf{V}\in\mathbb{C}^{N_sV\times N_sV}$ are the new matrices to be optimized. Specially, $\mathbf{\Lambda}=[\text{diag}\{\sqrt{p_1},\sqrt{p_2},\dots,\sqrt{p_{N_sV}}\};\boldsymbol{0}_{(D-N_sV)\times N_sV}]$, where the non-zero values are in descending order. According to \eqref{eq29}-\eqref{eq31}, the objective function of (P2) can be simplified as
\begin{align}\label{eq37}
     I(\mathbf{z}_{a};\mathbf{z}_{b})&=-\log_2\big(|\mathbf{I}-\mathbf{V}\mathbf{\Lambda}^2\mathbf{V}^H(\mathbf{V}\mathbf{\Lambda}^2\mathbf{V}^H+\mathbf{\Gamma}_a)^{-1}\nonumber\\
     &\mathbf{V}\mathbf{\Lambda}^2\mathbf{V}^H(\mathbf{V}\mathbf{\Lambda}^2\mathbf{V}^H+\mathbf{\Gamma}_b)^{-1}|\big),
\end{align}
where $\mathbf{\Lambda}^2=\mathbf{\Lambda}\mathbf{\Lambda}^T$.

Applying the Woodbury matrix inversion lemma, we further simplify \eqref{eq37} as a function of $\mathbf{\Lambda}$ which is present at the top of the next page. $(a)$ holds because the determinant of a unitary matrix equals to 1, i.e. $|\mathbf{V}|=1$.

\begin{figure*}[t] 
    \begin{align}\label{eq38}
    I(\mathbf{z}_{a};\mathbf{z}_{b})
    &=-\log_2\left(|\mathbf{I}-\hat{\sigma}^{-2}_a\mathbf{V}\mathbf{\Lambda}^2\mathbf{V}^H(\hat{\sigma}^{-2}_a\mathbf{V}\mathbf{\Lambda}^2\mathbf{V}^H+\mathbf{I}_a)^{-1}\times\hat{\sigma}^{-2}_b\mathbf{V}\mathbf{\Lambda}^2\mathbf{V}^H(\hat{\sigma}^{-2}_b\mathbf{V}\mathbf{\Lambda}^2\mathbf{V}^H+\mathbf{I}_b)^{-1}|\right) \nonumber\\ 
    &=-\log_2\left(|\mathbf{I}-(\mathbf{I}-(\mathbf{I}+\hat{\sigma}^{-2}_a\mathbf{V}\mathbf{\Lambda}^2\mathbf{V}^H)^{-1})\times(\mathbf{I}-(\mathbf{I}+\hat{\sigma}^{-2}_b\mathbf{V}\mathbf{\Lambda}^2\mathbf{V}^H)^{-1})|\right)\nonumber\\
    &\overset{(a)}{=}\log_2\left(|\mathbf{I}+\hat{\sigma}^{-2}_a\mathbf{\Lambda}^2|\right) + \log_2\left(|\mathbf{I}+\hat{\sigma}^{-2}_b\mathbf{\Lambda}^2|\right)-\log_2(\left|\mathbf{I}+\hat{\sigma}^{-2}_a\mathbf{\Lambda}^2 + \hat{\sigma}^{-2}_b\mathbf{\Lambda}^2)|\right).
    \end{align}
    \hrulefill
\end{figure*}
According to \eqref{eq38}, we have
\begin{align}
    I(\mathbf{z}_{a};\mathbf{z}_{b})=\sum_{i=1}^{D}\log_2\left(1+\frac{\hat{\sigma}^{-2}_a\hat{\sigma}^{-2}_bp_{i}^2}{(\hat{\sigma}^{-2}_a+\hat{\sigma}^{-2}_b)p_{i}+1}\right),  
\end{align} 
where $\hat{\sigma}_a^2=\frac{\sigma_a^2}{QP_b}$, $\hat{\sigma}_b^2=\frac{\sigma_b^2}{LP_a}$.

We substitute the \eqref{eq30} into the constraint of (P2) and get 
\begin{align}
    &\|\mathbf{W}\|^2_F=\Tr(\mathbf{W}\mathbf{W}^H) \nonumber\\
    &=\Tr\left((\mathbf{U}_h\mathbf{\Lambda}_h^{-H/2}\mathbf{U}\mathbf{\Lambda}\mathbf{V}^H)^*(\mathbf{U}_h\mathbf{\Lambda}_h^{-H/2}\mathbf{U}\mathbf{\Lambda}\mathbf{V}^H)^T\right)\nonumber\\
    &=\Tr(\mathbf{\Lambda}_h^{-1}\mathbf{U}^*\mathbf{\Lambda}\mathbf{\Lambda}^T\mathbf{U}^T)\nonumber\overset{(a)}{=}\Tr(\mathbf{\Lambda}_h^{-1}\mathbf{\Lambda}^2)=(M+1)^2N,
\end{align}
where $\mathbf{\Lambda}_h^{-1}=\mathbf{\Lambda}_h^{-T/2}\mathbf{\Lambda}_h^{-1/2}$ and (a) holds when $\mathbf{U}=\mathbf{I}_D$. Therefore, the constraint of (P2) is simplified as 
\begin{equation}
   \sum_{i=1}^{D}\frac{p_{i}}{p_{h,i}}=(M+1)^2N.
\end{equation}

\subsubsection{Design $p_i$}
After transforming $\mathbf{W}$ into $p_i$, the problem (P2) can be converted into (P3), which is given by
\begin{equation}\label{eq34}
\begin{aligned}
     \text{(P3)}~\mathop{\max_{p_{i}}}~&\sum_{i=1}^{D}\log_2\left(1+\frac{\hat{\sigma}^{-2}_a\hat{\sigma}^{-2}_bp_{i}^2}{(\hat{\sigma}^{-2}_a+\hat{\sigma}^{-2}_b)p_{i}+1}\right) \\
      &s.t.~\sum_{i=1}^{D}\frac{p_{i}}{p_{h,i}}=(M+1)^2N.
\end{aligned}
\end{equation}

We resort to finding the optimal value of the SKR by the Lagrangian multiplier solution. According to \cite{bib6}, the solution to a concave function over a convex solution set can be guaranteed to be a global maximum. Therefore, we first discuss the concavity of the objective function of (P3). We derive the first-order and second-order partial derivative of $I(\mathbf{z}_a;\mathbf{z}_b)$, which is shown at the top of the next page.
\begin{figure*}[t]
\begin{align}
       \frac{\partial I(\mathbf{z}_a;\mathbf{z}_b)}{\partial p_{i}}
       &=\frac{\hat{\sigma}^{-2}_a\hat{\sigma}^{-2}_b(\hat{\sigma}^{-2}_a\!+\!\hat{\sigma}^{-2}_b)p_{i}^2+2\hat{\sigma}^{-2}_a\hat{\sigma}^{-2}_bp_{i}}{\ln2\left(\hat{\sigma}^{-2}_a\hat{\sigma}^{-2}_bp_{i}^2 \!+\!(\hat{\sigma}^{-2}_a\!+\!\hat{\sigma}^{-2}_b)p_{i}+1\right)\left((\hat{\sigma}^{-2}_a\!+\!\hat{\sigma}^{-2}_b)p_{i}\!+\!1\right)},\\
        \frac{\partial^2 I(\mathbf{z}_a;\mathbf{z}_b)}{\partial p^2_{i}}
        &=\frac{2\hat{\sigma}^{-2}_a\hat{\sigma}^{-2}_b}
        {\ln2\left(\hat{\sigma}^{-2}_a\hat{\sigma}^{-2}_bp_{i}^2 + (\hat{\sigma}^{-2}_a+\hat{\sigma}^{-2}_b)p_{i}+1\right)}-\frac{2(\hat{\sigma}^{-2}_a+\hat{\sigma}^{-2}_b)\times\left(\hat{\sigma}^{-2}_a\hat{\sigma}^{-2}_b(\hat{\sigma}^{-2}_a\!+\!\hat{\sigma}^{-2}_b)p_{i}^2+2\hat{\sigma}^{-2}_a\hat{\sigma}^{-2}_bp_{i}\right)}{\ln2\left(\hat{\sigma}^{-2}_a\hat{\sigma}^{-2}_bp_{i}^2\!+\!(\hat{\sigma}^{-2}_a+\hat{\sigma}^{-2}_b)p_{i}\!+\!1\right)\left((\hat{\sigma}^{-2}_a\!+\!\hat{\sigma}^{-2}_b)p_{i}\!+\!1\right)^2} \nonumber\\
        &-\frac{(\hat{\sigma}^{-2}_a\hat{\sigma}^{-2}_b(\hat{\sigma}^{-2}_a+\hat{\sigma}^{-2}_b)p_{i}^2+2\hat{\sigma}^{-2}_a\hat{\sigma}^{-2}_bp_{i})^2}{\ln2(\hat{\sigma}^{-2}_a\hat{\sigma}^{-2}_bp_{i}^2\!+\!(\hat{\sigma}^{-2}_a\!\!\!+\!\hat{\sigma}^{-2}_b)p_{i}\!+\!1)^2((\hat{\sigma}^{-2}_a\!+\!\hat{\sigma}^{-2}_b)p_{i}\!+\!\!1)^2}.
\end{align}
\hrulefill
\end{figure*}

The first-order partial derivative of $I(\mathbf{z}_a;\mathbf{z}_b)$ is greater than zero, which means the decomposed $D$ functions is monotonically increasing. If the objective function is concave, the Hessian matrix should be semi-negative definite. Because $\frac{\partial^2 I(\mathbf{z}_a;\mathbf{z}_b)}{\partial p_{i}\partial p_{j}} = 0$, the objective is concave if $\frac{\partial^2 I(\mathbf{z}_a;\mathbf{z}_b)}{\partial p_{i}^2} \leq 0$. The second-order partial derivative function is greater than zero at the interval of $[0,p_{co}]$ and less than zero at the interval of $[p_{co},+\infty]$, where $p_{co}$ is the solution of $\frac{\partial^2 I(\mathbf{z}_a;\mathbf{z}_b)}{\partial p_{i}^2}=0$. That is, each decomposed function is convex at the interval of $[0,p_{co}]$ and concave at the interval of $[p_{co},+\infty]$. Therefore, we will find the global maximum if we constraint the power in the concave interval. 
From the above analysis, we find there are two challenges in solving the problem (P3). The first is to find the active set with non-zero power. The second is to consider the convex interval of each decomposed function. In order to simplify (P3), we will derive the Karush-Kuhn-Tucker (KKT) conditions of the (P3). According to the KKT conditions, we further propose a water-filling algorithm to find the optimal $p_i$, $i = 1,\dots,D$.

\subsubsection{Water Filling Algorithm} 
We first derive the KKT condition of the problem, and then we find the threshold of each decomposed function. If the power allocated to the $i$-th subchannel is greater than the threshold $\gamma_i=p_{h,i}p_{co}$, $i=1,\dots,D$, the subchannel is activated with non-zero power. Finally, we propose a water-filling algorithm to solve the problem with lower-bound constraints.

The Lagrangian function with respect to $p_{i}$ is given by
\begin{equation}
f = I(\mathbf{z}_a;\mathbf{z}_b) - \mu\left(\sum_{i=1}^D\frac{p_{i}}{p_{h,i}}-(M+1)^2N\right),
\end{equation}
where $\mu\geq0$ is the water-filling level.
The corresponding KKT conditions are
\begin{equation}\label{eq41}
\left\{
\begin{aligned}
 &~y_i(p_i)=p_{h,i}\frac{\partial f}{\partial p_{i}} = p_{h,i}\frac{\partial I(\mathbf{z}_a;\mathbf{z}_b)}{\partial p_{i}}=\mu, \\
 &~\mu\left( \sum_i\frac{p_{i}}{p_{h,i}}-(M+1)^2N \right)=0,\\
 &~\sum_i\frac{p_{i}}{p_{h,i}}\leq(M+1)^2N,
\end{aligned}
\right.
\end{equation}
where $y_i(p_i)$ is the increasing rate of the power allocated to the $i$-th subchannel.

Next, we transform the KKT condition into the water-filling algorithm and find the solution. According to \cite{bib5}, we define $\mathcal{M}_{inactive}=\{i|p_i\leq\gamma_i, i=1,\dots, D\}$ as the set of inactive cascaded channels, i.e. the allocated power is lower than the threshold. Also, define $\mathbb{I}_i$ as the indicator function. If $p_i\in\mathcal{M}_{inactive}$, $\mathbb{I}_i = 0$, otherwise $\mathbb{I}_i = 1$. Define $g_i(\mu)$ as the inverse function of $y_i(p_i)$. Therefore, the KKT condition \eqref{eq41} can be transformed into the following conditions.
\begin{equation}\label{eq42}
\left\{
\begin{aligned}
 &~p_i = g_i(\mu)\mathbb{I}_i + \gamma_i(1-\mathbb{I}_i),\\
 &~\sum_{i=1}^D g_i(\mu)\mathbb{I}_i + \gamma_i(1-\mathbb{I}_i) = P,\\
\end{aligned}
\right.
\end{equation}
where $P=(M+1)^2N$. 

We use the water-filling algorithm to solve the problem \eqref{eq42}, which is shown in Algorithm 1. In line 1, we initialize $\mathbb{I}_i=1$ for $i=1,\dots,D$, which means all subchannels are active in the initial phase. Specially, the $m$-th subchannel is the $m$-th element of $\mathbf{h}_r$. In line 2, according to the KKT conditions in \eqref{eq42}, we calculate the initial $p_i$ and the water level, $\mu$. In line 4, we find the set of $\mathcal{M}_{inactive}$. In line 5, we set $\mathbb{I}_i=0$ for $i\in\mathcal{M}_{inactive}$, which is the process to find the inactive subchannels and allocate power $\gamma_i$ to them. In line 6, according to $\mathbb{I}_i$, we calculate the $p_i$ and $\mu$. We repeat the steps from lines 4 to 6 until all $p_i\geq\gamma_i$. When the algorithm ends, the channels belonging to the active subchannel set are allocated with the optimal power, while those belonging to the inactive subchannel set are allocated with power $\gamma_i$.

\makeatletter
\newcommand{\removelatexerror}{\let\@latex@error\@gobble}
\makeatother

\begin{figure}[!t]
  \renewcommand{\algorithmicrequire}{\textbf{Input:}}
  \renewcommand{\algorithmicensure}{\textbf{Output:}}
  \removelatexerror
  \begin{algorithm}[H]
    \caption{Water-Filling Algorithm} \label{Al1}
    \begin{algorithmic}[1]
      \REQUIRE  $\{p_{h,i}\}$, $P$, $\mu_{max}$, $\mu_{min}$, $\hat{\sigma}_a$, $\hat{\sigma}_b$, $\epsilon_1$, $\epsilon_2$, $\gamma_i$;
      \ENSURE  $\{p_{i}\}$, $\{\mathbb{I}_i\}$.
      \STATE Initialize $\mathbb{I}_i=1$ for $i=1,\dots,D$; 
      \STATE {Calculate $\{p_i\}$ and $\mu$ through Algorithm 2;}
      \REPEAT 
      \STATE {Set $\mathcal{M}_{inactive}=\{i|p_i\leq\gamma_i, i=1,\dots, D\}$};
      \STATE {Substitute $\mathbb{I}_i=0$ for $i\in\mathcal{M}_{inactive}$;}
      \STATE {Calculate $\{p_i\}$ and $\mu$ through Algorithm 2;}
      \UNTIL {length(find $(p_i<\gamma_i)=0$).}
    \end{algorithmic}
  \end{algorithm}
\end{figure}

\begin{figure}[!t]
  \renewcommand{\algorithmicrequire}{\textbf{Input:}}
  \renewcommand{\algorithmicensure}{\textbf{Output:}}
  \removelatexerror
  \begin{algorithm}[H]
    \caption{Two-dimensional Bisection Algorithm} \label{Al2}
    \begin{algorithmic}[1]
      \REQUIRE $\{p_{h,i}\}$, $P$, $\mu_{max}$, $\mu_{min}$, $\hat{\sigma}_a$, $\hat{\sigma}_b$, $\epsilon_1$, $\epsilon_2$;
      \ENSURE  $\{p_{i}\}$, $\mu$.
      \STATE {Set $\mu = (\mu_{min}+\mu_{max})/2$;}
      \FOR{$i = 1, \dots, D$}
      \IF {$\mathbb{I}_i = 1$}
      \STATE {Do bisection search of $p_{i}$ to satisfy $|y_i-\mu|\leq \epsilon_1$;}
      \ELSE
      \STATE {$p_{i}=\gamma_i$;}
      \ENDIF
      \ENDFOR
      \REPEAT 
      \FOR{$i = 1, \dots, D$}
      \IF {$\sum\frac{p_{i}}{p_{h,i}}<P$}
      \STATE {$\mu_{max} = (\mu_{min}+\mu_{max})/2$;}
      \ELSE
      \STATE {$\mu_{min} = (\mu_{min}+\mu_{max})/2$;}
      \ENDIF
      \STATE {Set $\mu = (\mu_{min}+\mu_{max})/2$;}
      \IF{$\mathbb{I}_i = 1$}
      \STATE {Do bisection search of $p_{i}$ to satisfy $|y_i-\mu|\leq \epsilon_1;$}
      \ELSE
      \STATE {$p_{i}=\gamma_i$;}
      \ENDIF
      \ENDFOR
      \UNTIL {$|\sum\frac{p_{i}}{p_{h,i}}-P|\leq \epsilon_2$.}
    \end{algorithmic}
  \end{algorithm}
\end{figure}

In Algorithm 1, we should calculate $p_i$ and $\mu$ in each loop. The  $p_i$ is derived from \eqref{eq42}, where $g_i(\mu)$ is the core parameter. However, it is hard to find the closed-form expression of $g_i(\mu)$. Therefore, we introduce two-dimensional bisection search to calculate $p_i$ and $g_i(\mu)$, as shown in Algorithm 2. 
In line 1, we set the initial $\mu$ as $\mu=(\mu_{min}+\mu_{max})/2$. From line 2 to 8, for all subchannels, if $\mathbb{I}_i = 1$, we apply the bisection search to find the initial $p_i$ to meet the requirement of $|y_i-\mu|\leq \epsilon_1$; If $\mathbb{I}_i = 0$, we set $p_i = \gamma_i$. From line 10 to 22, we search the $p_i$ and $\mu$, and repeat it until $|\sum\frac{p_{i}}{p_{h,i}}-P|\leq \epsilon_2$. When Algorithm 2 ends, we get the final $p_i$ and $\mu$, which will be returned to Algorithm 1.

According to the Algorithms 1 and 2, we obtain the optimal $p_i$, $i=1,\dots,D$. Substituting $p_i$ into \eqref{eq30}, we find the optimal solution, $\mathbf{W}^{\dag}$, of the problem (P1). In the next section, based on $\mathbf{W}^{\dag}$, we will design an algorithm to find the optimal precoding and phase shift matrices.

\subsection{Precoding and Phase Shift Matrices Design}\label{sec:design}
In order to obtain the phase shift matrix, $\bar{\boldsymbol{\Phi}}_i$, and the precoding matrix, $\mathbf{P}$, we try to decompose the $\mathbf{W}^{\dag}$ to $\bar{\boldsymbol{\Phi}}\otimes\mathbf{P}$. However, in most cases, the upper bound is not approachable because $\mathbf{W}^{\dag}$ cannot be decomposed into a Kronecker product of a matrix with unit-magnitude entries and an F-norm-constrained complex matrix. Therefore, we obtain $\bar{\boldsymbol{\Phi}}$ and $\mathbf{P}$ through solving the following optimization problem.
\begin{equation}\label{ob43}
  \begin{aligned}
    \text{(P4)}~\mathop{\min_{\bar{\boldsymbol{\Phi}},\mathbf{P}}} &\|\bar{\boldsymbol{\Phi}}\otimes\mathbf{P}-\mathbf{W}^{\dag}\|_F^2 \\
    \mathrm{ s.t. }~&\eqref{c1}-\eqref{c3}.
  \end{aligned}
\end{equation}

Next, we propose an algorithm to design the precoding and phase shift matrices. The special structure of the constraints inspires us to find a fast way to design the phase-shift matrix. According to \cite{bib14,bib15}, we design the phase-shift matrix as the Hadamard reflection pattern. 

Given $\bar{\boldsymbol{\Phi}}$, we design the precoding matrix
\begin{equation}
  \begin{aligned}
    \text{(P5)}~\mathbf{P}^{\dag}=\mathop{\arg\min_{\mathbf{P}}} &\|\bar{\boldsymbol{\Phi}}\otimes\mathbf{P}-\mathbf{W}^{\dag}\|^2_F \\
    \mathrm{ s.t. }~& \mathbf{P}^H\mathbf{P}= \mathbf{I}_N.
  \end{aligned}
\end{equation}

Given the phase shift matrix, we propose an algorithm to design the precoding matrix. To address the Kronecker product, the objective function of (P5) can be transformed as $\sum_{m=1}^{M+1}\sum_{t=1}^{V}\|\mathbf{W}_{m,t}-\phi_{m,t}\mathbf{P}\|^2_F$, where $\mathbf{W}_{m,t}\in\mathbb{C}^{N\times N_s}$ is an element of
\begin{equation}
\begin{aligned}
 &\mathbf{W} = 
\begin{bmatrix}
&\mathbf{W}_{1,1}&\mathbf{W}_{1,2}&\dots&\mathbf{W}_{1,V}\\
& \vdots & \vdots &\ddots& \vdots\\
&\mathbf{W}_{M+1,1}&\mathbf{W}_{M+1,2}&\dots&\mathbf{W}_{M+1,V}
\end{bmatrix}.\\
\end{aligned}
\end{equation}

\begin{figure*}[t]

\begin{align}
    \|\bar{\boldsymbol{\Phi}}\otimes\mathbf{P}-\mathbf{W}^{\dag}\|^2_F &=\sum_{m=1}^{M+1}\sum_{t=1}^{V}\|\mathbf{W}_{m,t}-\phi_{m,t}\mathbf{P}\|^2_F \nonumber\\
    &=(M+1)V\|\mathbf{P}-\sum_{m=1}^{M+1}\sum_{t=1}^{V}\frac{\phi_{m,t}^*\mathbf{W}_{m,t}}{(M+1)V}\|^2_F+\sum_{m=1}^{M+1}\sum_{t=1}^{V}\|\mathbf{W}_{m,t}\|^2_F-\frac{1}{(M+1)V}\|\sum_{m=1}^{M+1}\sum_{t=1}^{V}\phi_{m,t}^*\mathbf{W}_{m,t}\|^2_F \nonumber\\
    &\geq(M+1)V\|\mathbf{P}-\sum_{m=1}^{M+1}\sum_{t=1}^{V}\frac{\phi_{m,t}^*\mathbf{W}_{m,t}}{(M+1)V}\|^2_F.
\end{align}
\hrulefill
\end{figure*}

The objective function of (P5) can be further expanded, which is shown at the top of the next page. Given $\bar{\mathbf{\Phi}}$, we design the precoding matrix by solving the following optimization problem.
\begin{equation}\label{ob34}
  \begin{aligned}
   \text{(P6)}~\mathop{\min_{\mathbf{P}}}~
    & \|\mathbf{P}-\sum_{m=1}^{M+1}\sum_{t=1}^{V}\frac{\phi_{m,t}^*\mathbf{W}_{m,t}}{(M+1)V}\|^2_F\\
    \mathrm{ s.t. }~& \mathbf{P}^H\mathbf{P}=\mathbf{I}_N.
  \end{aligned}
\end{equation}
Since the orthogonal constraint is non-convex, we resort to the Grassmann manifold toolbox to solve the above problem\cite{manopt}.

\section{Key Generation Protocol}\label{sec:protocol}
\subsection{Preprocessing}
After the channel probing process in Section \ref{sec:channel_probing}, BS and UE acquire a series of measurements, i.e., $\mathbf{z}_{a}{(t)}$ and $\mathbf{z}_{b}{(t)}$, $t=1,\dots,T_d$, where $T_d$ is the number of measurements.  The LoS component of the Rician fading is not suitable for key generation because the LoS component is determined by the distance between BS and UE~\cite{zhang2020new}. To address the issue, we need to wipe off the LoS component in the measurements and extract the NLoS components. We subtract the mean of the measurements and get $\mathbf{z}_{u}{(t)} = \mathbf{z}_{u}{(t)}-\frac{1}{T_d}\sum_{t=1}^{T_d}\mathbf{z}_{u}{(t)}$, $u\in\{a,b\}$. Since the difference between the path-loss effect of measurements from $\mathbf{z}_{u}{(t)}$ should be mitigated \cite{intro18,sim2}, we normalize the measurements as $\widehat{\mathbf{z}}_{u}(t)=\mathbf{z}_{u}(t)(\frac{1}{T_dD}\sum_{t=1}^{T_d}\|\mathbf{z}_{u}(t)\|_F^{2})^{\frac{1}{2}}$. After normalization, the measurements of BS and UE has zero mean and unit variance. 

\subsection{Quantization}
After preprocessing the measurements, BS and UE convert the analog channel measurements, $\widehat{\mathbf{z}}_{u}(t)$, to binary sequences. According to Algorithm~1 in~\cite{zhang2016experimental}, we apply a single-bit cumulative distribution function (CDF)-based quantization, $\mathcal{Q}(\cdot)$. Each user will carry out quantization independently, given as
\begin{align}
    \boldsymbol{K}_u = \mathcal{Q}(\widehat{\mathbf{z}}_{u}).
\end{align}

Since the noise causes disagreements between the bit sequences of BS and UE, we use bit disagreement rate (BDR) to quantify the difference between the bit sequences of BS and UE after quantization \cite{sim6}. We define the BDR as
\begin{align}
    \text{BDR} =\frac{\sum_{i=1}^{l_k} |\boldsymbol{K}_a(i)- \boldsymbol{K}_b(i)|}{l_k},
\end{align} 
where $\boldsymbol{K}_a(i)$ and $\boldsymbol{K}_b(i)$ are the bits generated from quantization and $l_k$ is the key length.


\section{Numerical Results}
In this section, numerical results are given to elaborate the performance of our proposed RIS-assisted key generation scheme. 

\subsection{Setup}
\subsubsection{Device Configuration}
The BS is located on the $x$-axis, $(0, 0, 1)$,  with antenna spacing $d_a=\lambda/2$ and $\lambda=0.1$~m. The RIS is parallel to $y-z$ plane, where the first reflecting element is located at $(39, 4.9, 4.9)$. The side length of an element is normalized by the wavelength, i.e., $d_r = \frac{d_r}{\lambda}$. The UE is located at $(39, 4.2, 5.4)$. The transmit powers of the BS and UE are set identically as $P_t = P_a = P_b$ dBm. All noise powers are  set as $\sigma^2 = \sigma_a^2=\sigma_b^2=-96$ dBm. 

\subsubsection{Channel Configuration}
The Rician factor is set as $K = K_{ar} = K_{br} = K_{ba} $ dB. The path-loss effect is modeled as $\beta_{uv} = \beta_0(\frac{d_{uv}}{d_0})^{-\epsilon_{uv}}$, $u,v\in\{a,b,r\}$, where $\epsilon_{uv}$ is the path-loss exponent, $\beta_0=-30$~dB denotes the path-loss effect at $d_0=1$ m and $d_{uv}$ is the link distance. The path-loss exponents of the BS-RIS, UE-RIS and UE-BS links are set as $\epsilon_{ar} = 2$, $\epsilon_{br} =2$ and $\epsilon_{ba} =3.67$, respectively.

\subsubsection{Considered Algorithms}
The proposed and existing algorithms are defined as follows:

\begin{enumerate}    
    \item \textbf{MA w/o RIS (non-optimized):} There are a BS with multiple antennas (without precoding) and a UE. The BS and UE measure the direct channel, $\mathbf{h}$, and extract the randomness from it. 
    
    \item \textbf{MA w/ RIS (non-optimized):} There are a BS with multiple antennas (without precoding), a UE and a RIS (phase shift vector not optimized). We configure the phase shift vector and precoding matrix as $\mathbf{v}= \mathbf{1}_M$ and $\mathbf{P}= \mathbf{I}_N$, respectively. The BS and UE extract the randomness from the LS estimation of the equivalent channel, $\mathbf{h}_{e}(\mathbf{v})$. 

    \item \textbf{SA w/ RIS (optimized):} This case is to design the phase shift vector, $\mathbf{v}$, in a single-antenna system, which is applied in \cite{intro10, intro12}. Given the optimal $\mathbf{v}$, the BS and UE extract the randomness from $\mathbf{h}_{e}(\mathbf{v})$. The optimal $\mathbf{v}$ is derived in Appendix A.
       
    \item \textbf{MA w/ RIS (optimized):} This case is to measure the cascaded channel and distill secret keys from it in a multiple-antenna system with a RIS. According to Algorithms 1 and 2, we configure the $\mathbf{W}$. This case illustrates the upper bound of the RIS-assisted key generation. We calculate the SKR extracted from the measurements of the cascaded channel, $\mathbf{W}^T\mathbf{h}_{r}$.
    
    \item \textbf{The proposed algorithm:} We configure the precoding matrix, $\mathbf{P}$, and phase-shift matrix, $\bar{\boldsymbol{\Phi}}$, based on the proposed algorithm. The BS and UE extract the randomness from the LS estimation of the cascaded channel, $(\bar{\mathbf{\Phi}}^T\otimes\mathbf{P}^T)\mathbf{h}_{r}$.
    
\end{enumerate}

\subsection{Results}
In all the figures, solid lines and markers denote numerical results and simulation results, respectively. In the proposed algorithm case, the numerical results of SKR are calculated from \eqref{ob}, where $\bar{\mathbf{\Phi}}$ and $\mathbf{P}$ are solved from the problem (P4). In the MA w/ RIS (optimized) case, the numerical results of SKR is calculated from \eqref{ob}, where $\bar{\mathbf{\Phi}}\otimes\mathbf{P}$ is replaced by $\mathbf{W}$. In the MA w/o RIS (non-optimized) case, the numerical results of SKR are calculated according to\cite{intro18}, where the channel coefficients of the multiple-antenna system are set as $\mathbf{h}$ in this paper. In the MA w/ RIS (non-optimized) case, the numerical results of SKR is extracted from the channel coefficients in \eqref{eq13}, where $\mathbf{v}= \mathbf{1}_M$ and $\mathbf{P}= \mathbf{I}_N$. In the SA w/ RIS (optimized) case, the numerical results of SKR are calculated based on \eqref{eq57} in Appendix A. In order to verify the numerical results, we carry out Monte Carlo simulations using Matlab and employ \emph{ITE} toolbox \cite{sim5} to calculate the mutual information of the measurements of BS and UE.

\subsubsection{Evaluation of SKR}
We evaluated SKR against transmit power, the number of reflecting elements, the side length of an element and the Rician factor.

\begin{figure}[t]
  \subfigure[]{
	\begin{minipage}[t]{0.46\linewidth}
		\centering
		\includegraphics[width = 1.64in]{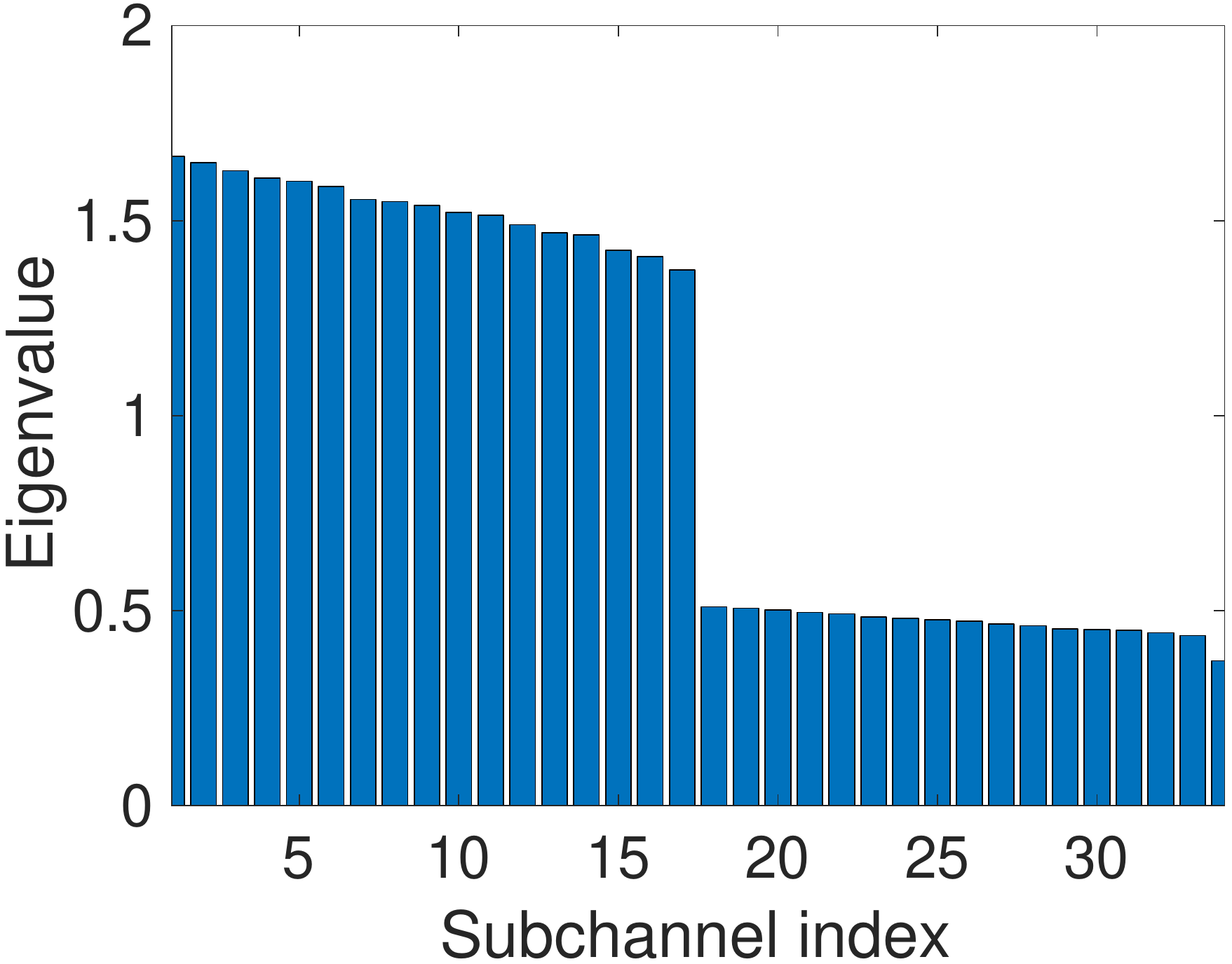}
	\end{minipage}
	}
   \subfigure[]{
	\begin{minipage}[t]{0.46\linewidth}
		\centering
		\includegraphics[width = 1.64in]{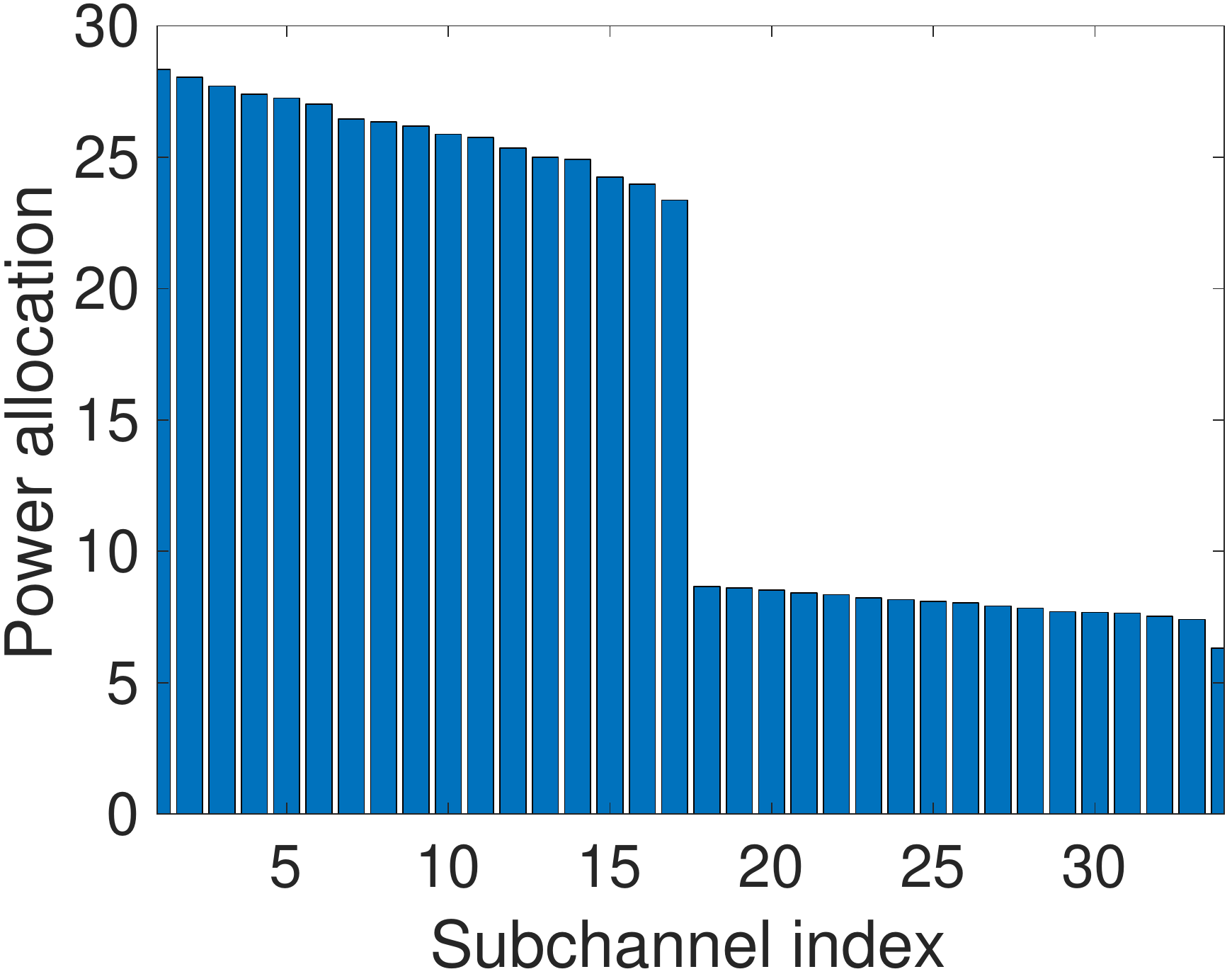}
	\end{minipage}
	}
\caption{(a) The eigenvalues of subchannels. (b) The power allocation for one channel realization. $P_t = 20$ dBm, $N=2$, $M=16$, $K = 10$ dB, $d_r = 0.5$.} 
\label{Figure1}
\end{figure}

Fig. \ref{Figure1} exhibits the eigenvalues of subchannels and power allocation results of Algorithm~1. The number of transmit antennas and reflecting elements are set as $N = 2$ and $M = 16$, respectively, so that $D = 34$. 
Fig. \ref{Figure1}(a) shows the eigenvalues of $34$ subchannels, which are in descending order. It is apparent that channel variance concentrates on seventeen subchannels whose eigenvalues are bigger than the others. Fig. \ref{Figure1}(b) shows the power allocation results of Algorithm 1. It can be seen that the power is mainly allocated to the first seventeen subchannels.

\begin{figure}[!t]
    \centering
    \includegraphics[width = 3.2in]{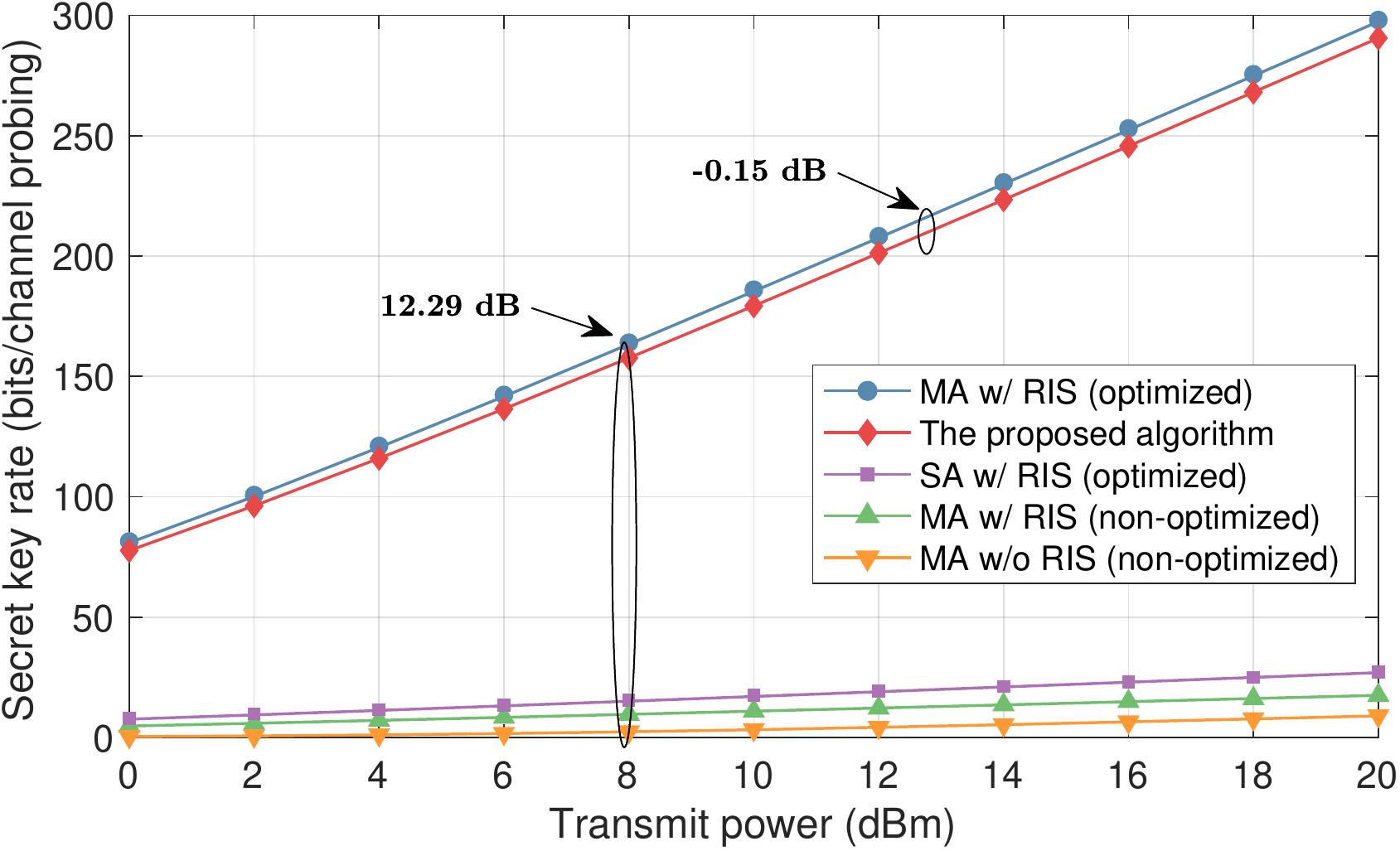}
    \caption{SKR versus transmit power. $N=2$, $M=16$, $K = 10$ dB, and $d_r = 0.5$.}
    \label{Fig2}
\end{figure}

Fig. \ref{Fig2} shows the SKR that BS and UE can achieve with each channel probing versus transmit power. 
The average gap between the SKR of MA w/ RIS (optimized) and MA w/o RIS (non-optimized) is about $12.29$ dB, which is a quite significant improvement. Compared to the SA w/ RIS (optimized) case, the proposed algorithm also has a great advantage, since the SKR extracted from the cascaded channel is greater than that extracted from the equivalent channel. Finally, we compare the SKR of the proposed algorithm and MA w/ RIS (optimized) cases. The average gap is about $-0.15$ dB, which means the SKR of the proposed algorithm case is approximately equivalent to the MA w/ RIS (optimized) case.

\begin{figure}[!t]
    \centering
    \includegraphics[width = 3.2in]{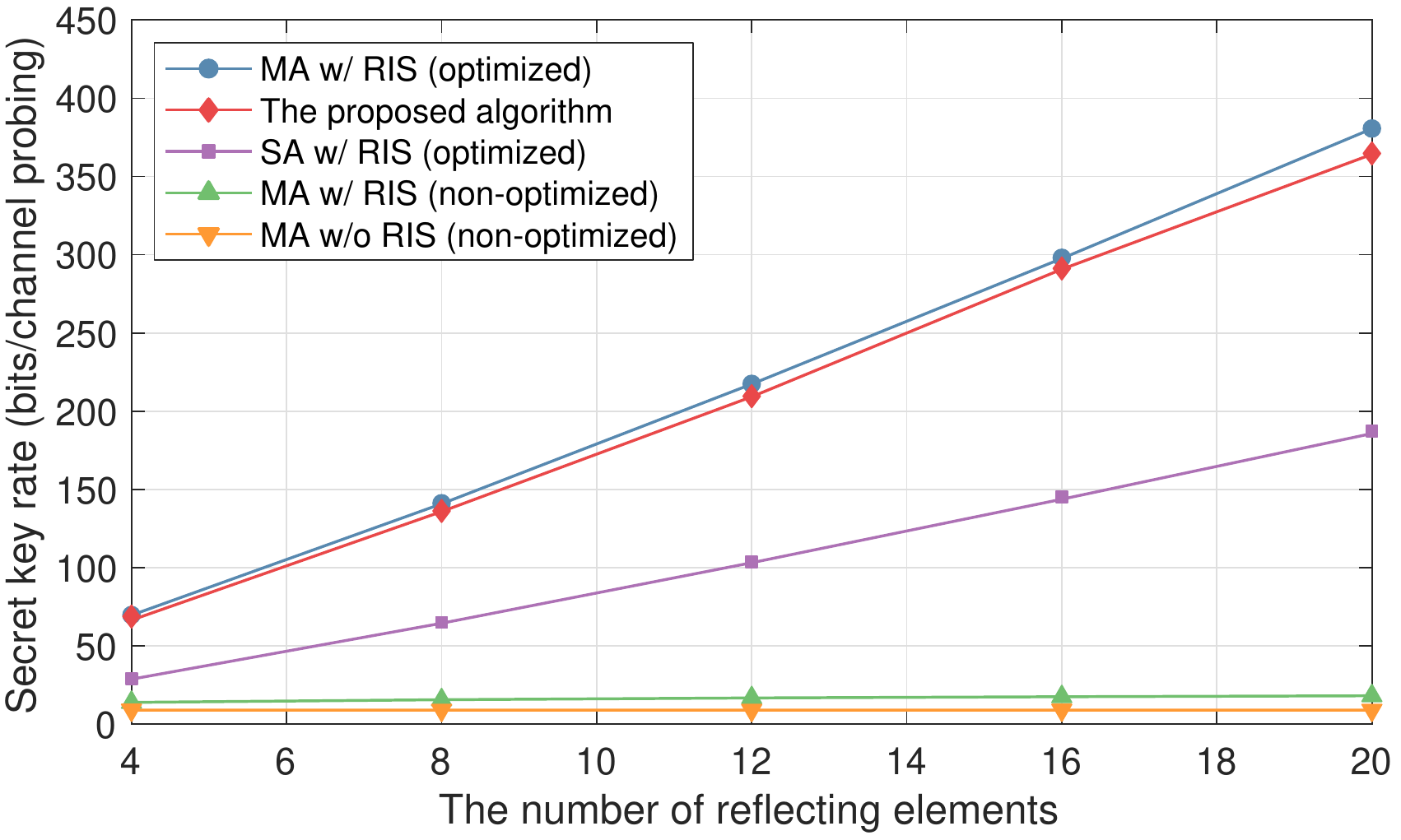}
    \caption{SKR versus the number of reflecting elements. $P_t = 20$ dBm, $N=2$, $K = 10$ dB, and $d_r = 0.5$.}
    \label{Fig4}
\end{figure}

Fig. \ref{Fig4} illustrates the SKR per channel realization for a different number of reflecting elements. It is observed that the SKR increases with the number of reflecting elements. Compared to the SA w/ RIS (optimized) case, the proposed algorithm exhibits considerable benefit, which means the increase of elements has a greater influence on the SKR extracted from the cascaded channel compared to the SKR extracted from the equivalent channel. What's more, the SKR of the MA w/o RIS (non-optimized) and MA w/ RIS (non-optimized) cases keep almost constant with the increase of the element numbers, which means element numbers cannot improve the SNR of these two cases greatly.

\begin{figure}[!t]
    \centering
    \includegraphics[width = 3.2in]{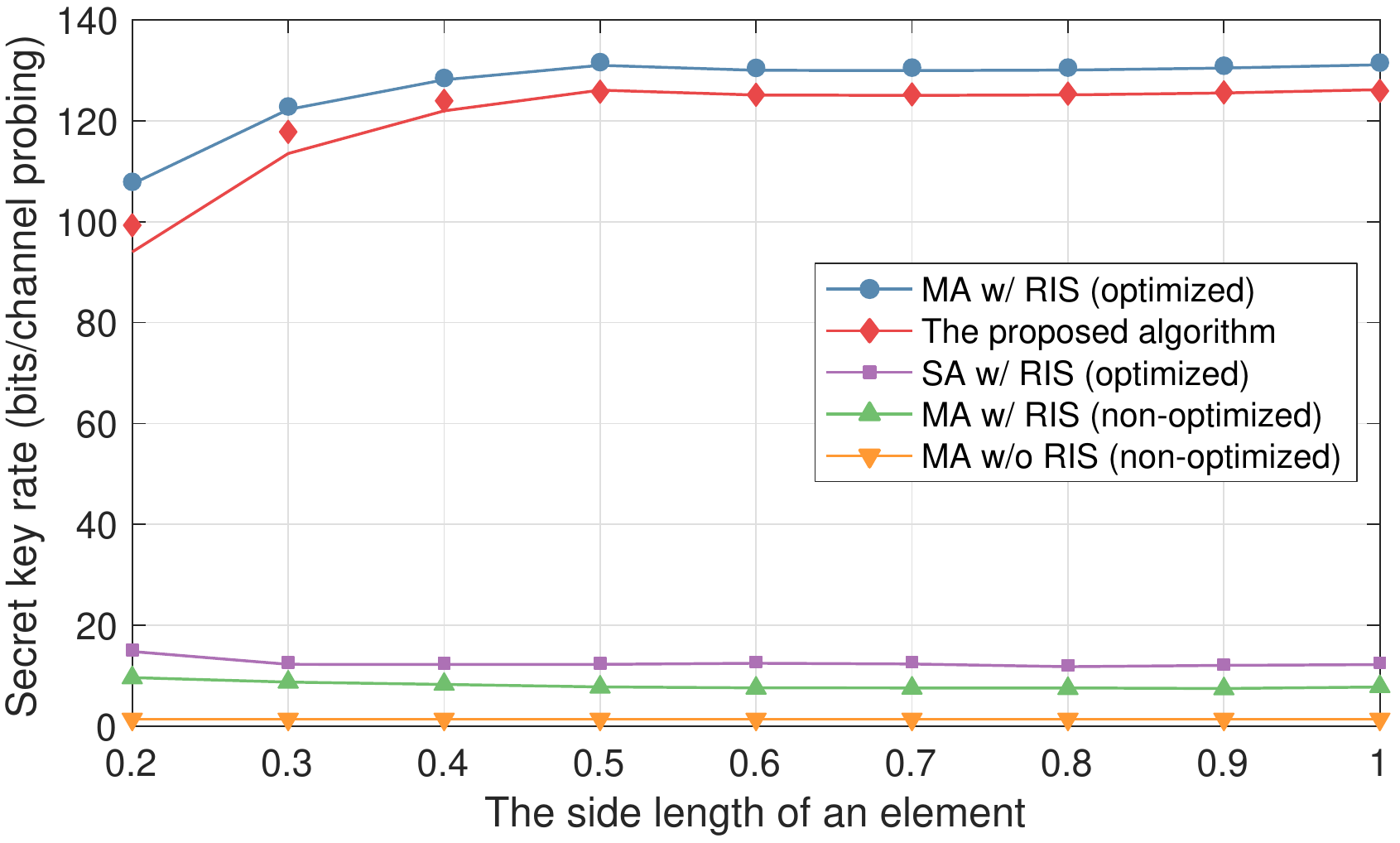}
    \caption{SKR versus the side length of an element. $P_t = 5$ dBm, $N = 2$, $M = 16$, $K = 10$ dB, and $d_r = 0.5$.}
    \label{Fig5}
\end{figure}

Fig. \ref{Fig5} investigates the impact of the side length of a reflecting element. The first observation is that the SKR increases with the side length of a reflecting element. As the side length gets larger from $0.2$ to $1$, the spatial correlation between subchannels gradually becomes lower and the subchannels eventually become uncorrelated. Therefore, the MA w/ RIS (optimized) and the proposed algorithm cases converge to a maximum value. What's more, there is a large performance gain in the proposed algorithm case compared with the SA w/ RIS (optimized) case. In the MA w/o RIS case, the SKR keeps constant because the SKR is determined by the direct channel. In the MA w/ RIS (non-optimized) and SA w/ RIS (optimized) cases, the SKR drops a little. We find that the quality of the equivalent channel gets worse with the increase of the side length.

\begin{figure}[!t]
    \centering
    \includegraphics[width = 3.2in]{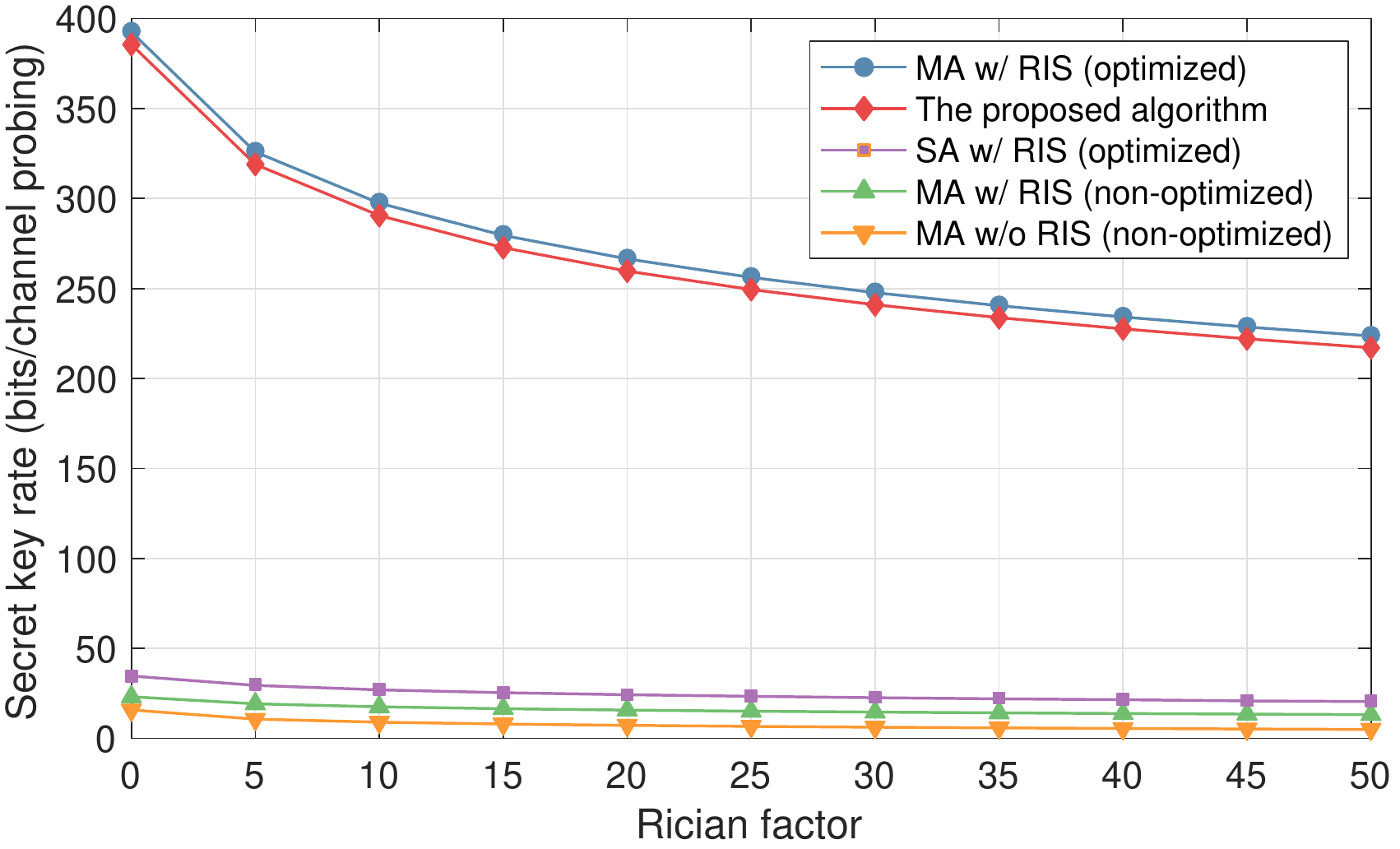}
    \caption{SKR versus Rician factor. $P_t = 20$ dBm, $N = 2$, $M = 16$, and $d_r = 0.5$.}
    \label{Fig6}
\end{figure}

Fig. \ref{Fig6} shows the SKR versus Rician factor $K$. As $K$ approaches infinity, the LoS component dominates. In contrast, the NLoS components dominate and the channel follows the Rayleigh distribution when $K$ approaches zero. It is apparent that SKR decreases with the Rician factor. When the Rician factor gets larger, the channel variance of the LoS fading becomes larger, which is not suitable for key generation. Compared to the MA w/ RIS (non-optimized) and SA w/ RIS (optimized) cases, our algorithm exhibits considerable improvement. 

\subsubsection{Evaluation of BDR and Randomness}
We also evaluated BDR and randomness after the channel measurements are converted to key bits.
The parameters of the system are set as $N = 2$, $M = 16$, $d_r = 0.5$, $K = 10$ dB. Here, we apply a single-bit CDF quantization to convert the measurements to bits. 

\begin{figure}[!t]
    \centering
    \includegraphics[width = 3.2in]{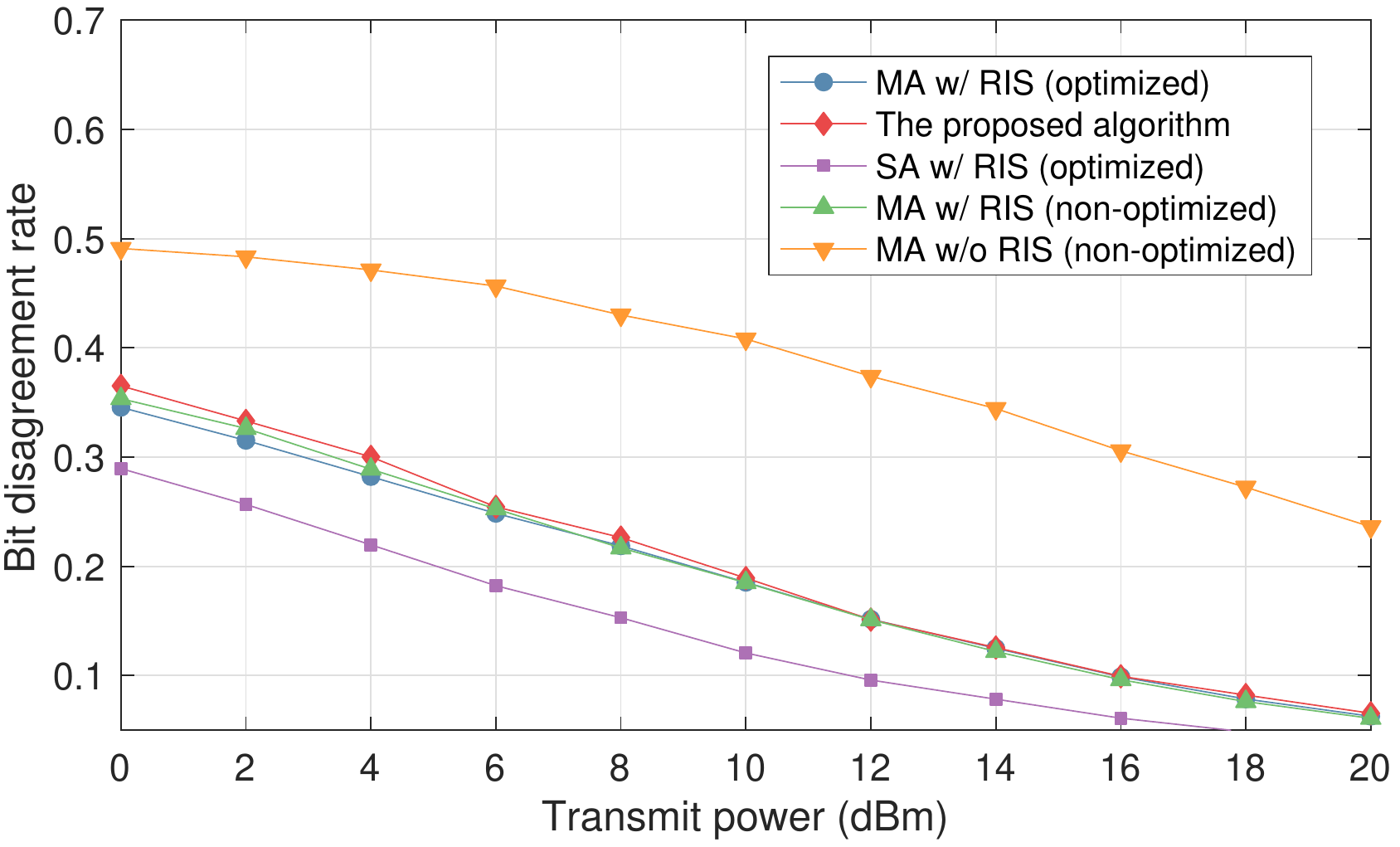}
    \caption{BDR versus transmit power. $N = 2$, $M = 16$, $d_r = 0.5$, $K = 10$~dB.}
    \label{Fig3}
\end{figure}

As shown in Fig. \ref{Fig3}, we show the BDR versus transmit power. Compared to the MA w/o RIS (non-optimized) case, the proposed algorithm greatly decreases the BDR. Since the equivalent channel is the combination of the direct and cascaded channels, the channel variance of the equivalent channel is bigger than the cascaded channel. Therefore, the BDR of the proposed algorithm is higher than the SA w/ RIS (optimized) case. 

The generated key should be random in order to meet the requirement of cryptographic applications~\cite{sim7}. We use the National Institute of Standards and Technology (NIST) test suite to validate the randomness, which is widely used in the key generation area \cite{sim8,sim9}. Each test returns a $p$ value. When the $p$ value is greater than $0.01$, the sequence passes the particular randomness test. We use the toolbox \cite{sim11} to perform 9 NIST statistical tests and the results are given in Table~\ref{tab:random}. All the results are greater than $0.01$, which means our algorithm is suitable for practical use. 
\begin{table}[t]
\caption{RANDOMNESS TEST RESULTS}
\centering
\begin{tabular}{|l|c|c|}
\hline
                        &MA w/ RIS (optimized) & The proposed algorithm\\ \hline
Frequency               & 0.61        & 0.72                                                                                         \\ \hline
Block frequency         & 0.24        & 0.91                                                                                         \\ \hline
Runs                    & 0.41        & 0.05                                                                                         \\ \hline
Longest run of 1s       & 0.77        & 0.12                                                                                         \\ \hline
DFT                     & 0.56        & 0.83                                                                                         \\ \hline
\multirow{2}{*}{Serial} & 0.06        & 0.12                                                                                         \\ \cline{2-3} 
                        & 0.07        & 0.32                                                                                         \\ \hline
Appro. entropy          & 0.99        & 0.017                                                                                        \\ \hline
Cum. sums. (fwd)        & 0.46        & 0.77                                                                                         \\ \hline
Cum. sums. (rev)        & 0.39        & 0.7                                                                                          \\ \hline
\end{tabular}
\label{tab:random}
\end{table}

\section{Conclusion}
This paper investigated a RIS-assisted key generation system, which jointly considered the design of the precoding and phase shift matrices to fully exploit the randomness from the cascaded channel. We first designed a water-filling algorithm to find the upper bound on the SKR of the system. Furthermore, we proposed an algorithm to obtain the phase shift and precoding matrices, which ensures the SKR approaches the upper bound. We found that the SKR is determined by the transmit power, the number of reflecting elements, the side length of an element and the Rician factor. Simulations validated that our protocol obtained a higher SKR than the existing algorithms.


%


\appendices
\section{Design of phase shift vector for the effective channel}

According to \eqref{equation3}, the equivalent channel is
\begin{align}
       \mathbf{h}_e(\mathbf{v}) =
       \begin{bmatrix}
       \mathbf{h}~ \mathbf{G}^T\text{diag}(\mathbf{f})
       \end{bmatrix}
        \bar{\mathbf{v}}.
\end{align}
If the BS is equipped with a single antenna, the equivalent channel is simplified as
\begin{align}
    h_e(\mathbf{v}) = 
      \begin{bmatrix}
       h~ \mathbf{g}^T\text{diag}(\mathbf{f})
       \end{bmatrix}
        \bar{\mathbf{v}},
\end{align}
where $h$, $\mathbf{g}\in\mathbb{C}^{M\times 1}$ and $\mathbf{f}\in\mathbb{C}^{M\times 1}$ are the UE-BS, BS-RIS and UE-RIS channels, respectively. Assume BS has the estimation noise $n_a$, $n_a\sim\mathcal{CN}(0,\sigma_a^2)$, and UE has the estimation noise $n_b$, $n_b\sim\mathcal{CN}(0,\sigma_b^2)$. Define $\widehat{h}_{e,a}(\mathbf{v})$ and $\widehat{h}_{e,b}(\mathbf{v})$ as the measurements of BS and UE, respectively. The covariance of the variable $[\widehat{h}_{e,a}(\mathbf{v}), \widehat{h}_{e,b}(\mathbf{v})]$ is given by
\begin{align}
    \mathbf{\Sigma} 
       &= \begin{bmatrix}
         p_e + \sigma_a^2~  &p_e \\
         p_e ~  &p_e+\sigma_b^2 
       \end{bmatrix},
\end{align}
where $p_e=\mathbb{E}\{h_e(\mathbf{v}) h_e(\mathbf{v})^*\}$ is the channel variance. If the direct and cascaded channels are uncorrelated, $p_e$ is given by
\begin{align}
    p_e =\bar{\mathbf{v}}^T\mathbf{R}_e\bar{\mathbf{v}}^*=\bar{\mathbf{v}}^T
    \begin{bmatrix}
      \sigma_h^2~&\mathbf{0}_M^T\\
      \mathbf{0}_M~&\mathbf{R}_{arb}
    \end{bmatrix}
    \bar{\mathbf{v}}^*,
\end{align}
where $\mathbf{R}_e$ is the channel covariance of the equivalent channel and $\mathbf{R}_{arb}=\mathbb{E}\{\text{diag}(\mathbf{f}^*)\mathbf{g}^*\mathbf{g}^T\text{diag}(\mathbf{f})\}$.

According to the differential entropy of a circularly symmetric Gaussian variable, the SKR is given by
\begin{align} \label{eq57}
    I(\widehat{h}_{e,a}(\mathbf{v}); \widehat{h}_{e,b}(\mathbf{v}))  
    &=\log_2\left(\frac{(p_e + \sigma_a^2)(p_e + \sigma_b^2)}{\mathbf{\Sigma}}\right) \nonumber\\
    &=\log_2\left(1+\frac{p_e}{\sigma_a^2+\sigma_b^2+\frac{\sigma_a^2\sigma_b^2}{p_e}}\right).
\end{align}
The SKR increases with $p_e$. To find the optimal SKR, we should solve the problem of $\mathop{\max_{\bar{\mathbf{v}}}}~\bar{\mathbf{v}}^T\mathbf{R}_e\bar{\mathbf{v}}$, where the objective function is a quadratic form. If there is a norm constraint, the optimal value is the biggest eigenvalue of $\mathbf{R}_e$ and the $\bar{\mathbf{v}}$ is the corresponding eigenvector. However, the reflection coefficient has the constraint of $|\bar{\phi}_{m}|=1$. Note that $\bar{\mathbf{v}}^T\mathbf{R}_e\bar{\mathbf{v}}^*=\Tr(\mathbf{R}_e\bar{\mathbf{v}}^*\bar{\mathbf{v}}^T)$. Define $\mathbf{\Theta}=\bar{\mathbf{v}}^*\bar{\mathbf{v}}^T$ and reformulate the problem as a semidefinite programming (SDP) problem. We should meet the constraints of $\mathbf{\Theta}\succeq 0$ and $\text{rank}(\mathbf{\Theta})=1$. Therefore, we have the following problem.

\begin{equation}
  \begin{aligned}
    \mathop{\max_{\mathbf{\Theta}}}~
    &\Tr\left[\mathbf{R}_e\mathbf{\Theta}\right]\\
    \mathrm{ s.t. }~& \mathbf{\Theta}_{i,i} = 1,~i=1,\dots,M+1,\\
     ~&\text{rank}(\mathbf{\Theta})=1.
  \end{aligned}
\end{equation}
Since the rank-one constraint is non-convex, we relax it and solve the above problem by the CVX\cite{cvx}. However, it may not generate a rank-one solution, i.e., rank$(\mathbf{\Theta})\neq1$. Furthermore, we apply the Gaussian randomization to obtain a feasible solution\cite{bib7}.

\ifCLASSOPTIONcaptionsoff
  \newpage
\fi



\bibliographystyle{IEEEtran}
\bibliography{IEEEabrv,RISMISO}
%

%








\end{document}